\documentclass[twocolumn,amsmath,amssymb,a4paper]{revtex4-1}

\usepackage{booktabs,tabu,multirow,dcolumn,tabularx,siunitx}
\usepackage{graphicx}
\usepackage{subcaption}
\usepackage{dcolumn}
\usepackage{bm}
\usepackage{rotating}

\usepackage[utf8]{inputenc}
\usepackage[english]{babel}
\usepackage[flushleft]{threeparttable}
\usepackage{bbm}
\usepackage{verbatim}
\usepackage{afterpage}

\usepackage{color}
\definecolor{lightblue}{rgb}{0.2,0.2,0.7}
\definecolor{darkblue}{rgb}{0,0.25,0.5}
\definecolor{redbrown}{rgb}{0.875,0.25,0.125}
\definecolor{darkgreen}{rgb}{0,0.5,0}
\definecolor{giuseppe}{rgb}{0.25,0.5,0.75}

\newcommand{\bra}[1]{\ensuremath{\langle #1 \vert}}
\newcommand{\ket}[1]{\ensuremath{\vert #1  \rangle}}
\newcommand{\braket}[2]{\ensuremath{\langle  #1 \vert #2  \rangle}}
\renewcommand{\b}[1]{\ensuremath{\mathbf{#1}}}
\renewcommand{\H}{\ensuremath{\text{H}}}

\newcommand{\lr}{\ensuremath{\text{lr}}}
\newcommand{\sr}{\ensuremath{\text{sr}}}

\newcommand{\RSH}{\ensuremath{\text{RSH}}}

\newcommand{\HF}{\ensuremath{\text{HF}}}

\begin{document}

\title{Range-separated double-hybrid density-functional theory applied to periodic systems}

\author{Giuseppe Sansone$^{1}$}
\author{Bartolomeo Civalleri$^1$}
\author{Denis Usvyat$^2$}
\author{Julien Toulouse$^{3,4}$}
\author{Kamal Sharkas$^{5}$}
\author{Lorenzo Maschio$^1$}\email{lorenzo.maschio@unito.it}
\affiliation{
$^1$Dipartimento di Chimica and NIS (Nanostructured Interfaces and Surfaces) Centre, Universit{\`a} di Torino, via Giuria 5, I-10125 Torino, Italy\\
$^2$ Institute for Physical and Theoretical Chemistry, Universit\"at Regensburg, Universit\"atsstrasse 31, D-93040 Regensburg (Germany)\\
$^3$Sorbonne Universit\'es, UPMC Univ Paris 06, UMR 7616, Laboratoire de Chimie Th\'eorique, F-75005 Paris, France\\
$^4$CNRS, UMR 7616, Laboratoire de Chimie Th\'eorique, F-75005 Paris, France\\
$^5$Department of Chemistry, State University of New York at Buffalo, Buffalo, New York 14260-3000, United States}

\date{June 12, 2015}

\begin{abstract}
Quantum chemistry methods exploiting density-functional approximations for short-range electron-electron interactions and second-order M{\o}ller-Plesset (MP2) perturbation theory for long-range electron-electron interactions have been implemented for periodic systems using Gaussian-type basis functions and the local correlation framework. The performance of these range-separated double hybrids has been benchmarked on a significant set of systems including rare-gas, molecular, ionic, and covalent crystals. The use of spin-component-scaled MP2 for the long-range part has been tested as well. The results show that the value of $\mu=0.5$ bohr$^{-1}$ for the range-separation parameter usually used for molecular systems is also a reasonable choice for solids. Overall, these range-separated double hybrids provide a good accuracy for binding energies using basis sets of moderate sizes such as cc-pVDZ and aug-cc-pVDZ.
\end {abstract}
\maketitle

\section{Introduction}

Among the wide variety of quantum chemistry methods proposed in the last
decade, a great fascination resides in the possibility to combine 
density-functional theory (DFT)\cite{HohKoh-PR-64,KohSha-PR-65} and explicit many-body
correlation methods, such as second-order M{\o}ller-Plesset
(MP2) perturbation theory,\cite{MolPle-PR-34} random-phase approximations (RPA)\cite{Scuseria:2008}
and coupled-cluster theory.\cite{RevModPhys.79.291} The hope is to get the best
out of both worlds, that is to combine a proper description of long-range
dispersion interactions (without introduction of empirical
corrections\cite{grimme2010}) and a good description of short-range electron correlations with a reduced dependence on the basis set.

Different ways to combine DFT and wave-function techniques have been
proposed. These include global double-hybrid approaches\cite{Gri-JCP-06,ShaTouSav-JCP-11} and range-separated approaches.\cite{Sav-INC-96,TouColSav-PRA-04,AngGerSavTou-PRA-05,GerAng-CPL-05b,GolWerSto-PCCP-05,GolWerStoLeiGorSav-CP-06,TouGerJanSavAng-PRL-09,JanHenScu-JCP-09}
In this work we focused on the latter type, which was
initially introduced by Stoll and Savin in the
80s.\cite{StoSav-INC-85} It has been shown that for molecular
complexes (and especially for dimers involving rare-gas atoms) it provides
excellent performance as regards bond lengths, dissociation energies, and harmonic
frequencies. In many cases, results are, quote, {``superior, with medium-size basis sets, to pure DFT and pure coupled-cluster calculations''}.\cite{GolWerStoLeiGorSav-CP-06}

Concerning the study of solids, we note that the introduction of wave-function-based
correlation treatment for periodic systems is relatively recent.\cite{PisSchCasUsvMasLorErb-PCCP-12,usvyat2013, gruneis2010,gruneis2011,booth2012,gruneis2013, vandevondele2012,delben2013,vandevondele_water}
Martinez-Casado and coworkers proposed to estimate the correlation energy as a MP2 contribution using a B3LYP reference state, and applied this scheme to the study of the adsorption of helium atoms on a MgO surface.\cite{martinez-casado2011-cc}
Del Ben and coworkers\cite{vandevondele2012} benchmarked the performance
of some global double hybrids on a set of molecular crystals. In particular, these authors
used the DSD-BLYP functional,\cite{Kozuch} that includes a spin-component-scaled (SCS)\cite{Gri-JCP-03} MP2 contribution.
Recently, some of us applied one-parameter double-hybrid methods to molecular crystals.\cite{Sharkas14}

In this work, we tested the performance of range-separated double-hybrid methods,~\cite{AngGerSavTou-PRA-05}
combining short-range density-functional approximations with long-range MP2 correlation, for
evaluating the cohesive energy of crystalline periodic systems. We 
implemented this approach in the {\sc Crystal}\cite{Dovesi2014} and {\sc
  Cryscor}\cite{PisSchCasUsvMasLorErb-PCCP-12} programs, that use a basis set
of Gaussian-type orbitals centred on atoms.
The MP2 correlation part is treated within a local approach, where the
pair-specific virtual space is spanned by projected atomic orbitals (PAOs), restricted to the
so-called domains, i.e. several atoms surrounding the considered localized occupied orbitals. This technique was initially proposed by Pulay in the
80s,\cite{pulay1983,PulSae-TCA-86} and extended by Werner, Sch\"{u}tz and
coworkers to high-level correlated methods for molecules.\cite{shw1999,sw2001,Schutz:2002p17093,Schutz:2002p18659,Werner:2003p154,Schutz:2003p18579,Schutz:2004p157,Adler:2009p48766,ws2011,laplace_lmp2,Tatiana_LocRev,kats2014}
The adaptation of the local MP2 method to periodic systems has been done in the
last decade, 
\cite{PisBusCapCasDovMasZicSch-JCP-05,pisani2008,PisSchCasUsvMasLorErb-PCCP-12}
and over the years has been successfully applied to quite a rich variety of systems.\cite{halo2009a,halo2011b,erba2011a,erba2011c,maschio2010b,martinez-casado2011,Tanskanen2012,GeF2,Ar_MgO,Si_cage_Ar,Constantinescu2013,He_MgO_bound,Hammerschmidt_2015}

The paper is structured as follows. In Section \ref{sec:methods}, a review of
the formal aspects of the methods and their extension to periodic systems is given. Section
\ref{sec:computational} contains the details on the model systems, basis sets
and computational parameters used in the calculations. The tests
of the range-separated double hybrids on three representative systems with the aim of studying the dependence on the
range-separation parameter are discussed in Section
\ref{sec:mu}. Further, Section \ref{sec:results} presents the
benchmarks of several range-separated double-hybrid approximations on a wider set of
systems. Finally, conclusions and perspectives on future work are provided in
Section \ref{sec:conclusions}. In Appendix \ref{appendix:dimers}, the methods employed in this study have been additionally tested on molecular dimers cut out from the bulk systems.

\section{Computational methods}
\label{sec:methods}

\subsection{Periodic range-separated hybrid scheme}
\label{sec:rs}
In the range-separated hybrid (RSH) scheme,~\cite{AngGerSavTou-PRA-05} the ground-state energy is approximated with the following minimization over (normalized) single-determinant wave functions $\Phi$
\begin{eqnarray}
E_\text{RSH}=
    \min_{\Phi}\left\{
       \bra{\Phi} \hat{T}+\hat{V}_{\text{ext}}+\hat{W}_{\text{ee}}^\text{\text{lr}} \ket{\Phi}
      +E_{\text{Hxc}}^\text{\text{sr}} \left[ n_{\Phi} \right]
               \right\},
\label{eq:RSHmin}
\end{eqnarray} 
where $\hat{T}$ is the kinetic energy operator, $\hat{V}_{\text{ext}}$ is the external potential (nuclei-electron + nuclei-nuclei interactions) operator, $\hat{W}_{\text{ee}}^\text{lr}$ is a long-range electron-electron interaction operator made with the long-range interaction $w_{\text{ee}}^\text{lr}(r)=\text{erf}(\mu r)/r$, and $E_{\text{Hxc}}^\text{sr}[n_\Phi]$ is the corresponding $\mu$-dependent short-range Hartree-exchange-correlation density functional evaluated at the density of $\Phi$. The parameter $\mu$ controls the range of the separation. It is somewhat clearer to rewrite Eq.~(\ref{eq:RSHmin}) as
\begin{eqnarray}
E_\text{RSH} &=& \min_{\Phi} \Bigl\{ \bra{\Phi} \hat{T}+\hat{V}_{\text{ext}} \ket{\Phi}
      +E_{\text{H}}[n_\Phi] + E_{\text{x},\HF}^{\lr}[\Phi]
\nonumber\\
&&  + E_{\text{xc}}^\sr \left[ n_{\Phi} \right] \Bigl\},
\label{eq:RSHmin2}
\end{eqnarray} 
where $E_{\text{H}}[n_\Phi]$ is the usual Hartree energy (with the Coulomb electron-electron interaction $w_\text{ee}(r)=1/r$), $E_{\text{x},\HF}^{\lr}[\Phi]$ is the long-range Hartree-Fock (HF) exchange energy, and $E_{\text{xc}}^\sr [n_{\Phi}]$ is the short-range exchange-correlation energy. The minimization in Eq.~(\ref{eq:RSHmin2}) leads to familiar hybrid Kohn-Sham (KS)-type self-consistent equations determining the RSH orbitals $\phi_i$ and orbital energies $\varepsilon_i$
\begin{eqnarray}
    \hat{F}^\RSH \ket{\phi_i} = \varepsilon_i \ket{\phi_i},
\end{eqnarray} 
with the RSH Fock operator 
\begin{eqnarray}
\hat{F}^\RSH = \hat{T}+\hat{V}_{\text{ext}}+\hat{V}_\H + \hat{V}_{\text{x},\HF}^{\lr} +\hat{V}_{\text{xc}}^{\sr},
\end{eqnarray} 
where $\hat{V}_\H$ is the usual Hartree potential operator, $\hat{V}_{\text{x},\HF}^{\lr}$ is the long-range HF exchange potential operator, and $\hat{V}_{\text{xc}}^{\sr}$ is the short-range exchange-correlation potential operator. For $\mu=0$ the RSH scheme reduces to pure KS DFT, while for $\mu\to\infty$ it reduces to pure HF theory.

The RSH scheme and other similar range-separated hybrid DFT scheme are available in many molecular quantum chemistry programs. A very similar scheme to RSH, called RSHX,~\cite{GerAng-CPL-05a} where the separation is done on the exchange energy only, has been implemented for periodic systems using a plane-wave/projector-augmented-wave (PAW) approach.~\cite{GerAngMarKre-JCP-07} 
{ Another kind of range-separated hybrids, called screened-exchange hybrids~\cite{HeyScuErn-JCP-03} which use short-range HF exchange instead of long-range HF exchange, has also been implemented for periodic systems using Gaussian-type basis functions~\cite{HeyScu-JCP-04b} or a PAW approach~\cite{PaiMarHumKreGerAng-JCP-06}.}
 Here, we give the main equations for a spin-restricted closed-shell RSH scheme for periodic systems using local basis functions (see, e.g., Refs.~\onlinecite{PisDov-IJQC-80,TowZupCau-CPC-96,KudScu-CPL-98,KudScu-PRB-00} for more details on periodic HF or KS implementations with local basis functions).
 Due to translational symmetry, the crystalline orbitals are labeled by a wave vector $\b{k}$ and expanded as $\ket{\phi_i(\b{k})} = \sum_{\mu} c_{\mu i}(\b{k}) \ket{\psi_\mu({\b{k})}}$ where $\ket{\psi_\mu({\b{k})}} = N^{-1/2}\sum_{\b{g}} e^{i \b{k}\cdot\b{g}} \ket{\chi_{\mu}^\b{g}}$ are Bloch functions, and $N$ is the number of crystal cells and $\braket{\b{r}}{\chi_{\mu}^\b{g}}=\chi_{\mu}(\b{r}-\b{g})$ is an atomic-orbital basis function (a contraction of Gaussian-type orbital functions) located in the cell characterized by the direct lattice vector $\b{g}$. The orbital coefficients and energies are found by solving the self-consistent-field (SCF) equation at each point $\b{k}$
\begin{eqnarray}
\b{F}^\RSH(\b{k}) \b{c}_{i}(\b{k}) = \varepsilon_i(\b{k}) \b{S}(\b{k}) \b{c}_{i}(\b{k}),
\end{eqnarray}
with the RSH Fock matrix $F_{\mu \nu}^\RSH(\b{k}) =\sum_\b{g} e^{i \b{k} \cdot\b{g}} F_{\mu \nu \b{g}}^\RSH$ where $F_{\mu \nu \b{g}}^\RSH = \bra{\chi_{\mu}^\b{0}}  \hat{F}^\RSH \ket{\chi_{\nu}^\b{g}}$, and similarly for the overlap matrix $S_{\mu \nu}(\b{k})$. The matrix $F_{\mu \nu \b{g}}^\RSH$ is expressed as
\begin{eqnarray}
F^\RSH_{\mu \nu \b{g}} = h_{\mu \nu \b{g}} + J_{\mu \nu \b{g}}+ K^\lr_{\mu \nu \b{g}} + V^{\sr}_{\text{xc},\mu \nu \b{g}},
\end{eqnarray} 
where $h_{\mu \nu \b{g}} = \bra{\chi_{\mu}^{\b{0}}}  \hat{T} + \hat{V}_{\text{ext}} \ket{\chi_{\nu}^{\b{g}}}$ are the kinetic + external potential integrals, $V^{\sr}_{\text{xc},\mu \nu \b{g}} = \bra{\chi_{\mu}^{\b{0}}}  \hat{V}^\sr_{\text{xc}} \ket{\chi_{\nu}^{\b{g}}}$ are the short-range exchange-correlation potential integrals, $J_{\mu \nu \b{g}}$ is the usual Hartree potential contribution calculated with Coulombic two-electron integrals
\begin{eqnarray}
J_{\mu \nu \b{g}} = \sum_{\lambda \sigma \b{m} \b{l}} P_{\lambda \sigma\b{l}} \; (\chi_{\mu}^\b{0} \chi_{\nu}^\b{g} | w_\text{ee} | \chi_{\sigma}^{\b{m}} \chi_{\lambda}^\b{m+l}),
\end{eqnarray} 
and $K^\lr_{\mu \nu \b{g}}$ is the long-range HF exchange potential contribution calculated with long-range two-electron integrals
\begin{eqnarray}
K^\lr_{\mu \nu \b{g}} = -\frac{1}{2}\sum_{\lambda \sigma \b{m} \b{l}} P_{\lambda \sigma\b{l}} \; (\chi_{\mu}^\b{0} \chi_{\lambda}^\b{m+l} | w_\text{ee}^\lr | \chi_{\sigma}^{\b{m}} \chi_{\nu}^\b{g}).
\end{eqnarray} 
In these expressions, the density matrix $P_{\lambda \sigma\b{l}}$ is obtained from the occupied orbital coefficients as
\begin{eqnarray}
P_{\lambda \sigma\b{l}} = \frac{2}{v} \int_{\text{BZ}} \sum_i c_{\lambda i}(\b{k}) c_{\sigma i}^*(\b{k}) \theta(\varepsilon_\text{F} - \varepsilon_i(\b{k})) e^{i \b{k}\cdot\b{l}} \text{d}\b{k},
\end{eqnarray} 
where $v$ is the volume of the Brillouin zone (BZ) and $\varepsilon_\text{F}$ is the Fermi energy.

Finally, the RSH energy per unit cell takes the form
\begin{eqnarray}
E_\RSH = \sum_{\mu \nu \b{g}} P_{\nu \mu\b{g}} \left[ h_{\mu \nu \b{g}} + \frac{1}{2} \left( J_{\mu \nu \b{g}} + K^\lr_{\mu \nu \b{g}} \right) \right] + E_\text{xc}^\sr,\;
\end{eqnarray} 
where the short-range exchange-correlation energy per unit cell is, e.g. for generalized-gradient approximations,
\begin{eqnarray}
E_\text{xc}^\sr = \int_\text{unit cell} n(\b{r}) \epsilon_{\text{xc}}^{\sr}(n(\b{r}),\bm{\nabla} n(\b{r})) \; \text{d}\b{r},
\end{eqnarray} 
where the integration is over one unit cell and $n(\b{r}) = \sum_{\lambda \sigma \b{m} \b{l}}  P_{\lambda \sigma\b{l}} \; \chi^\b{m+l}_\lambda(\b{r}) \chi^\b{m}_\sigma(\b{r})$ is the electron density. In this work, we use either the short-range local-density-approximation (LDA) exchange-correlation functional of Ref.~\onlinecite{PazMorGorBac-PRB-06} or the short-range Perdew-Burke-Ernzerhof (PBE) exchange-correlation functional of Ref.~\onlinecite{GolWerStoLeiGorSav-CP-06} (which is a modified version of the one of Ref.~\onlinecite{TouColSav-JCP-05}). The method will thus be referred to as RSHLDA or RSHPBE, respectively.

\subsection{Periodic long-range local second-order M{\o}ller-Plesset correction}

The RSH scheme does not contain long-range correlation, but it can be used as a reference for a nonlinear Rayleigh-Schr\"odinger perturbation theory~\cite{AngGerSavTou-PRA-05,FroJen-PRA-08,Ang-PRA-08} to calculate the long-range correlation energy. At second order, the long-range correlation energy is rigorously given by a standard MP2 expression evaluated with RSH orbitals and orbital energies, and long-range two-electron integrals~\cite{AngGerSavTou-PRA-05,GerAng-CPL-05b}. Here, we give the main equations of the long-range local MP2 correction for periodic systems.

After the periodic RSH calculation, the crystalline RSH canonical occupied orbitals are transformed into localized~\cite{ZicDovSau-JCP-05} symmetry-adapted~\cite{CasZicPia-TCA-06} mutually orthogonal Wannier functions (WFs). As regards the virtual orbital space, mutually nonorthogonal PAOs are constructed by projecting the individual atomic-orbital basis functions on the virtual space~\cite{PulSae-TCA-86}. The long-range first-order double-excitation amplitudes $T^\lr_{\b{i}\b{a},\b{j}\b{b}}$ are then obtained by iteratively solving the following system of linear equations~\cite{PulSae-TCA-86,PisBusCapCasDovMasZicSch-JCP-05,PisMasCasHalSchUsv-JCC-08}
\begin{align} 
K^{\lr}_{\b{i}\b{a},\b{j}\b{b}} &+ \sum_{(\b c \b d) \in [\b i \b j]} \Bigg \{ F^\RSH_{\b a \b c} \; T^\lr_{\b{i}\b{c},\b{j}\b{d}} \; S_{\b d \b b} + S_{\b a \b c} \; T^\lr_{\b{i}\b{c},\b{j}\b{d}} \; F^\RSH_{\b d \b b} \Bigg \}
\nonumber\\ 
&- \sum_{(\b c \b d) \in u[\b j]} S_{\b a \b c}  \sum_{\b k \,{\text{near}}\, \b j} F^\RSH_{\b i \b k} \; T^\lr_{\b{k}\b{c},\b{j}\b{d}} \; S_{\b d \b b} 
\nonumber\\ 
&- \sum_{(\b c \b d) \in u[\b i]} S_{\b a \b c} \sum_{\b k \,{\text{near}}\, \b i}  T^\lr_{\b{i}\b{c},\b{k}\b{d}} \; F^\RSH_{\b k \b j} \; S_{\b d \b b}  =0,
\label{Amp}
\end{align}
where $\b{i}$, $\b{j}$, $\b{k}$ refer to WF occupied orbitals, and $\b{a}$, $\b{b}$, $\b{c}$, $\b{d}$ to PAO virtual orbitals (bold indices combine the index within the unit cell and the lattice vector). The locality is exploited by restricting the sums over PAO pairs $(\b c \b d)$ to the pair domain $[\b i \b j]$ of PAOs spatially close to at least one of the WF $\b i$ or $\b j$, or to the domain $u[\b i]$ (or $u[\b j]$) which is the union of all $[\b i \b k]$ (or $[\b j \b k]$) where the sum over $\b k$ is in turn limited to WFs spatially close to $\b i$ (or $\b j$). In Eq.~(\ref{Amp}), $K^{\lr}_{\b{i}\b{a},\b{j}\b{b}}=(\b i \b a| w_\text{ee}^\lr|\b j \b b)$ are the long-range two-electron exchange integrals in the WF/PAO basis, $S_{\b a \b b}$ is the overlap between PAOs, and $F^\RSH_{\b i \b j}$ and $F^\RSH_{\b a \b b}$ are elements of the RSH Fock matrix in WF and PAO basis, which is obtained by transformation of the Fock matrix in the atomic-orbital basis $F^\RSH_{\mu \nu \b{g}}$.

The $K^{\lr}_{\b{i}\b{a},\b{j}\b{b}}$ integrals are efficiently evaluated through a robust\cite{Dunlap:2000p1595} density-fitting scheme, suitably adapted for periodic systems.\cite{maschio2007,Usvyat:2007p10,MasUsv-PRB-08,schutz2010} 
By introducing an auxiliary basis set of Gaussian-type functions -- here indicated by indices ${\b P}$ and ${\b Q}$ -- the integrals are approximated as
 \begin{eqnarray}
K^{\lr}_{\b{i}\b{a},\b{j}\b{b}}&\approx&
   \sum_{\b P} d_{\b i \b a}^{\b P} (\b P | w_\text{ee}^\lr | \b j \b b) 
+ \sum_{\b Q} (\b i \b a | w_\text{ee}^\lr | \b Q ) d_{\b j \b b}^{\b Q} \nonumber \\
&-& \sum_{\b P \b Q}  d_{\b i \b a}^{\b P}  (\b P| w_\text{ee}^\lr  |\b Q) d_{\b j \b b}^{\b Q},
\label{df1}
\end{eqnarray}
with fitting coefficients
\begin{equation}
d_{\b i \b a}^{\b P}= \sum_{\b Q}({\b i \b a}| 1/r |\b Q) \left[J^{-1} \right]_{{\b Q}, {\b P}}  \quad,
\label{df2}
\end{equation}
where $\left[J^{-1} \right]_{{\b Q}, {\b P}}$  is the ${\b Q}, {\b P}$ element of the inverse of the matrix of Coulomb integrals over the auxiliary functions $J_{{\b P}, {\b Q}} = (\b P| 1/r |\b Q)$.
In Eqs. (\ref{df1}) and (\ref{df2}), the summation over fitting functions ${\b P}$ and ${\b Q}$ is limited to suitable local fitting domains.

The long-range local MP2 correlation energy per unit cell is then given as
\begin{eqnarray}
E_\text{c,MP2}^{\lr} = \sum_{(\b i \b j) \in P} \sum_{(\b a \b b) \in [\b i \b j]} K^{\lr}_{\b{i}\b{a},\b{j}\b{b}} \; (2 \; T^{\lr}_{\b{i}\b{a},\b{j}\b{b}}-T^{\lr}_{\b{i}\b{b},\b{j}\b{a}}),
\label{EcMP2PerCell}
\end{eqnarray} 
where the first sum is over occupied WF pairs $(\b i\b j)$ taken from a truncated list $P$, in which the first WF $\b i$ is located in the reference unit cell and the second WF $\b j$ is restricted within a given distance to the first WF $\b i$. The method obtained after adding the long-range local MP2 correlation energy to the RSH energy will be referred to as RSHLDA+MP2 or RSHPBE+MP2. Obviously, for $\mu=0$, the method reduces to pure LDA~\cite{PerWan-PRB-92} or pure PBE~\cite{PerBurErn-PRL-96}, while for $\mu=\infty$ it reduces to pure MP2.
{
We note that, since exactly the same range-separated Fock operator used in the SCF iterations is adopted for the evaluation of the long-range local MP2 correlation energy, no contribution from single excitations~\cite{usvyat2010} arises. }

We also consider the SCS variant~\cite{Gri-JCP-03} of MP2
\begin{eqnarray}
E_\text{c,SCS-MP2}^{\lr} &=& \sum_{(\b i \b j) \in P} \sum_{(\b a \b b) \in [\b i \b j]} K^{\lr}_{\b{i}\b{a},\b{j}\b{b}} \; \Bigl[(c_{\text{OS}}+c_\text{SS}) \; T^{\lr}_{\b{i}\b{a},\b{j}\b{b}}
\nonumber\\
&&- c_\text{SS} \; T^{\lr}_{\b{i}\b{b},\b{j}\b{a}} \Bigl],
\label{EcMP2PerCell}
\end{eqnarray} 
where the opposite-spin (OS) and same-spin (SS) coefficients, taken from the original work~\cite{Gri-JCP-03}, are $c_{\text{OS}} = 6/5$ and $c_{\text{SS}} = 1/3$. For molecular systems, the SCS variant of MP2 has been shown to significantly improve the accuracy of MP2 for energy differences and properties. For compactness of notation, the method will be referred to as RSHLDA+SCS or RSHPBE+SCS.

\begin{figure}
        \centering
        \begin{subfigure}[c]{0.2\textwidth}
                \includegraphics[width=2cm,height=2cm,keepaspectratio]{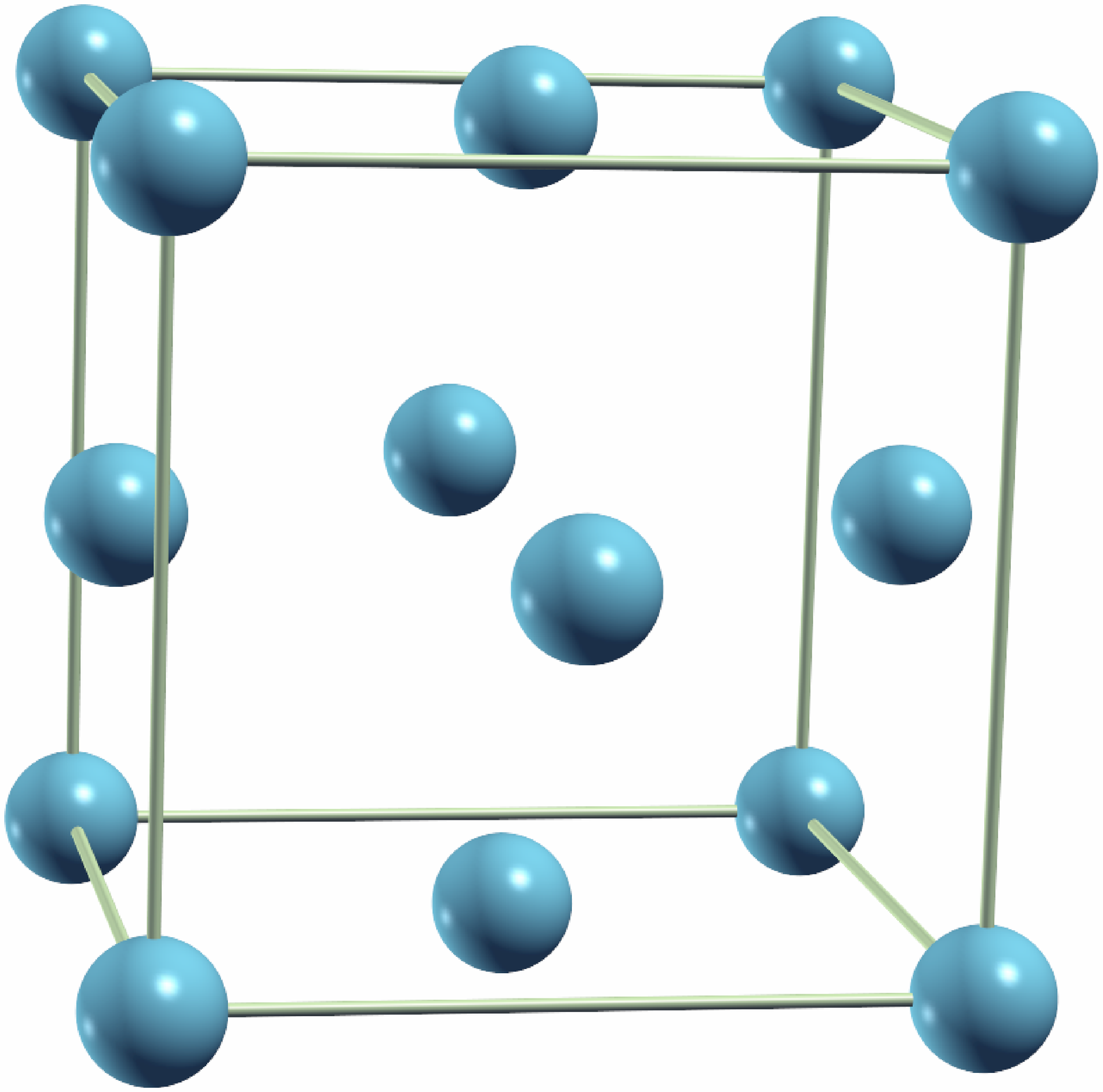}
                \caption{Neon}
        \end{subfigure}
        \qquad
        \begin{subfigure}[c]{0.2\textwidth}
                \includegraphics[width=2cm,height=2cm,keepaspectratio]{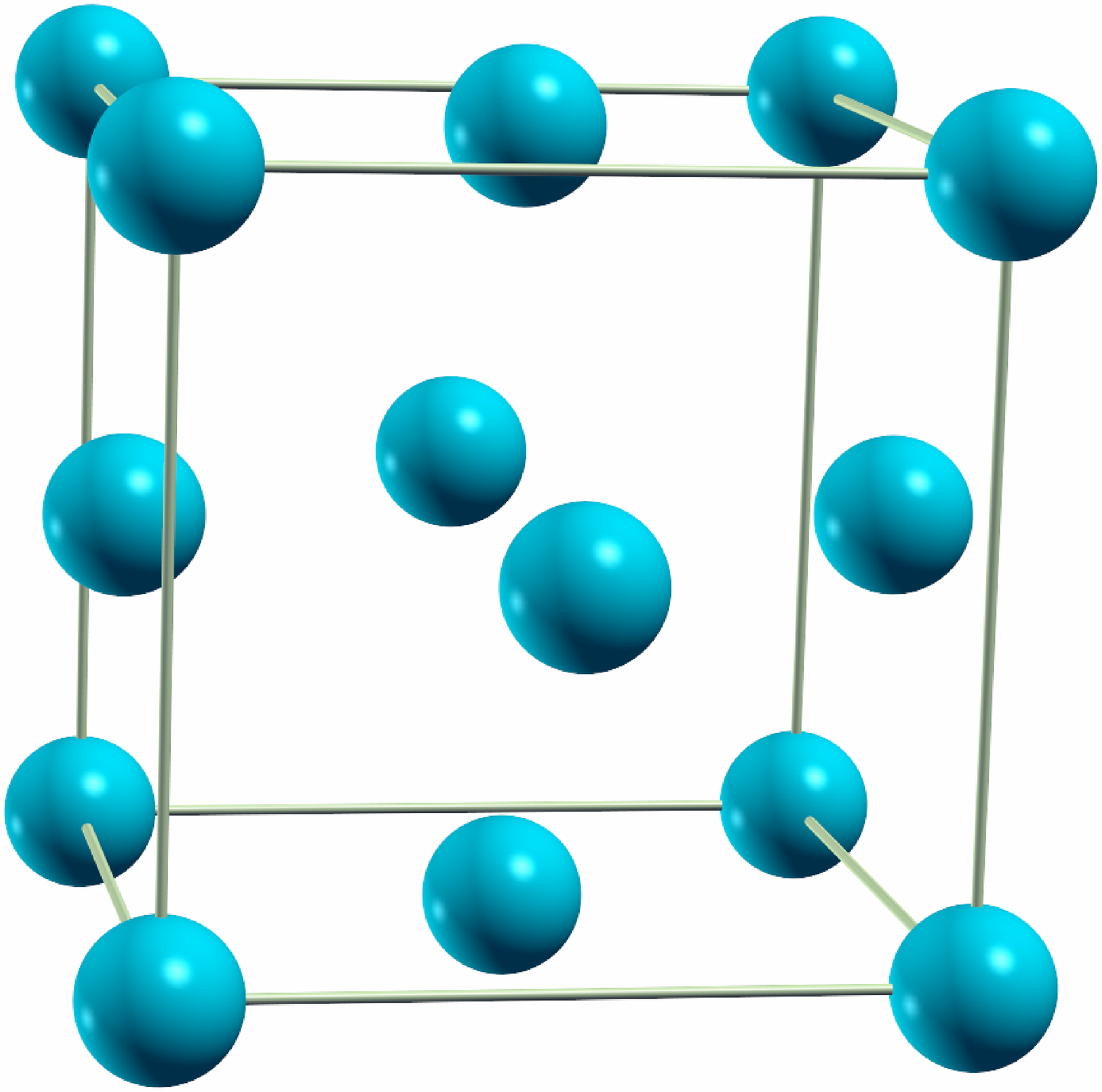}
                \caption{Argon}
        \end{subfigure}
\\
        \begin{subfigure}[c]{0.2\textwidth}
                \includegraphics[width=2cm,height=2cm,keepaspectratio]{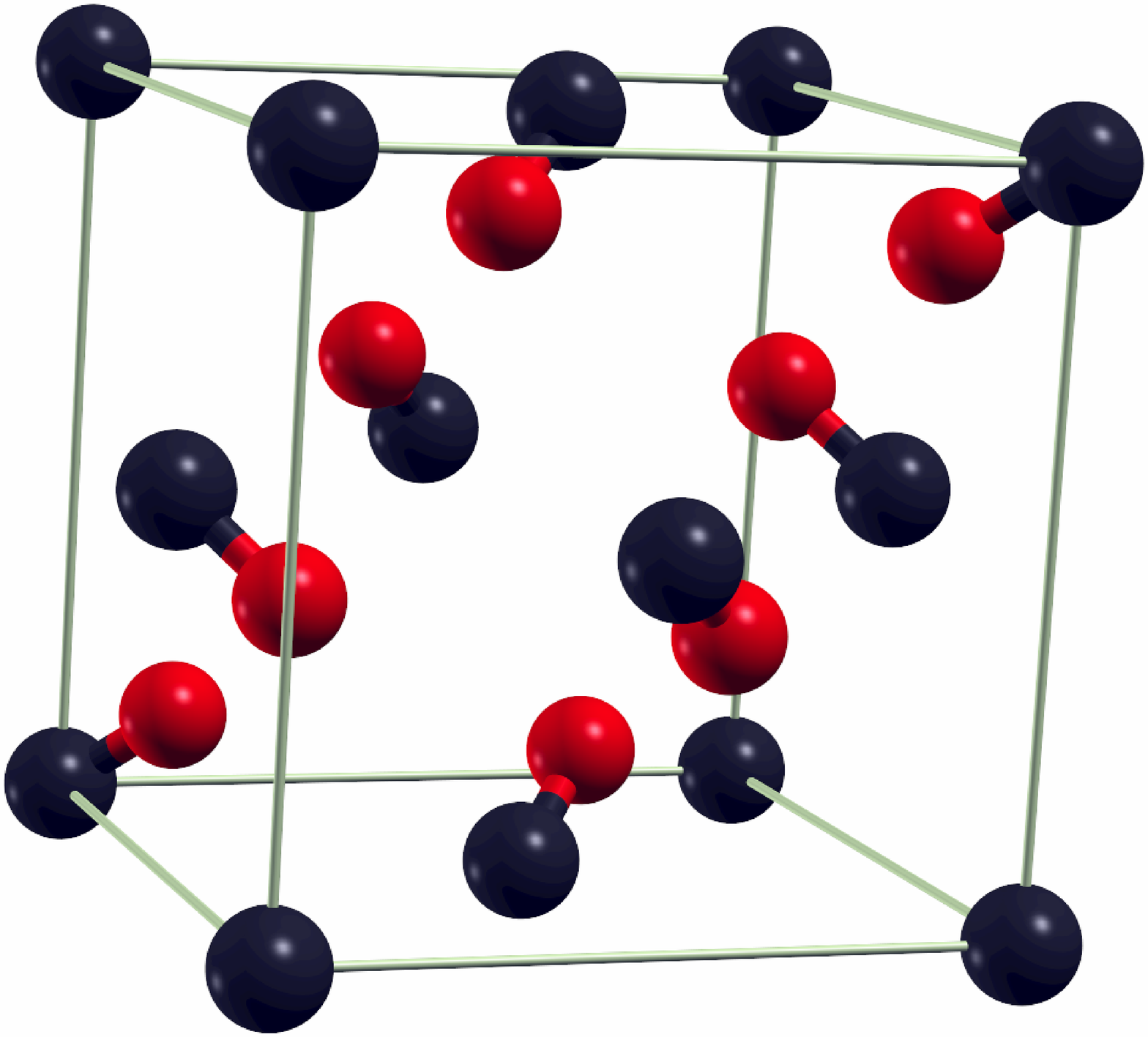}
                \caption{Carbon dioxide}
        \end{subfigure}
        \qquad
        \begin{subfigure}[c]{0.2\textwidth}
                \includegraphics[width=2cm,height=2cm,keepaspectratio]{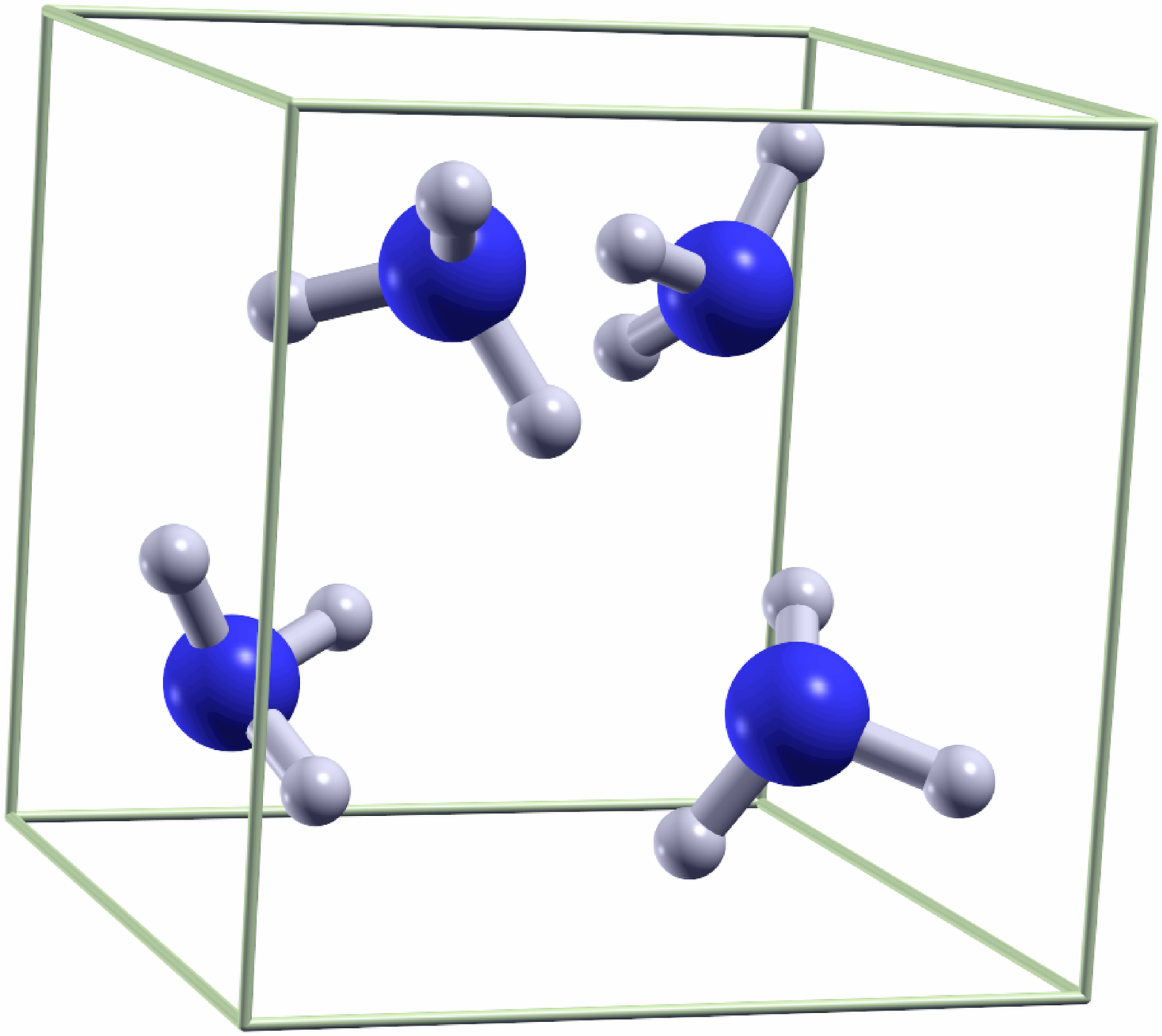}
                \caption{Ammonia}
        \end{subfigure}
        \qquad
        \begin{subfigure}[c]{0.2\textwidth}
                \includegraphics[width=2cm,height=2cm,keepaspectratio]{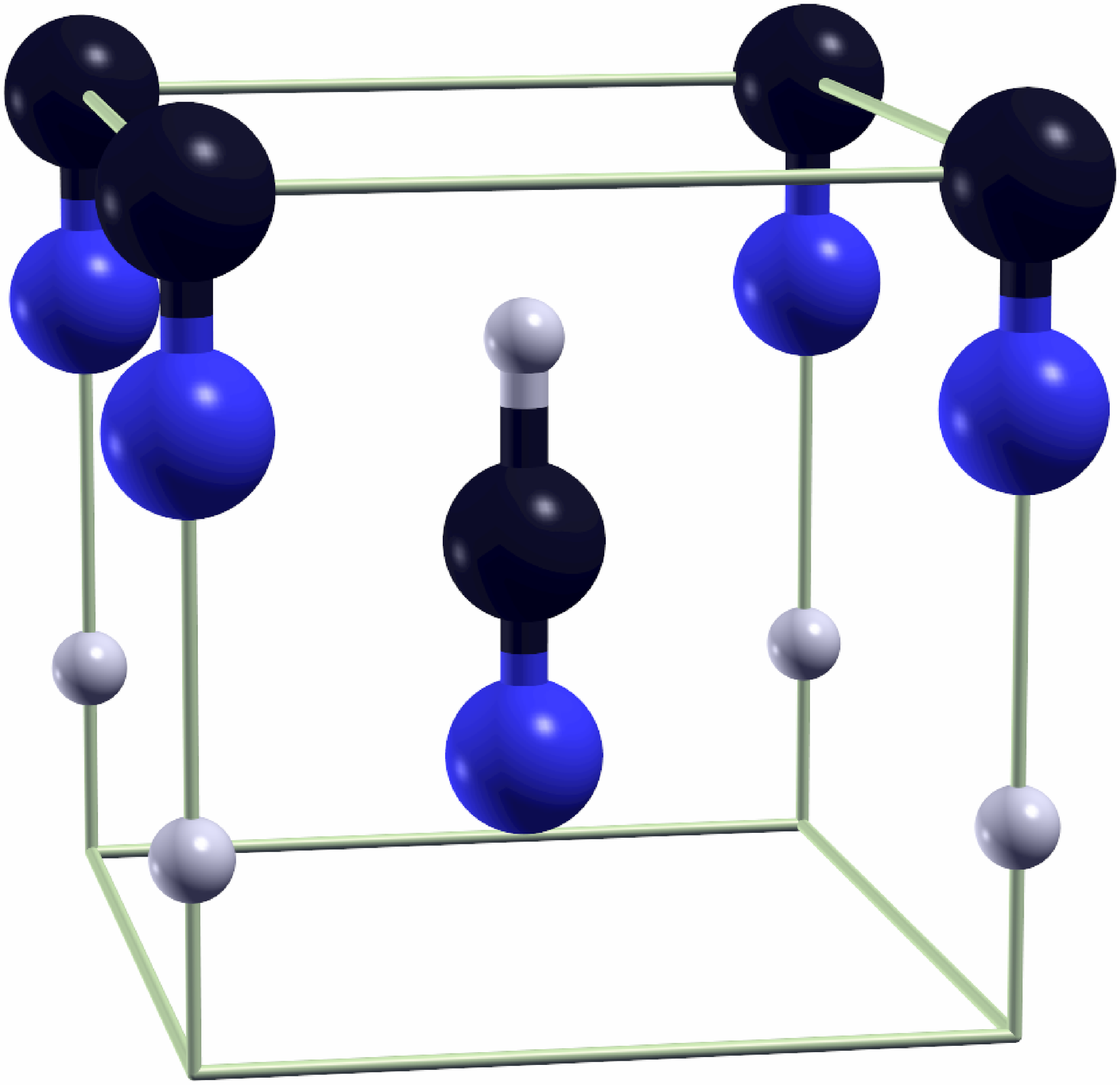}
                \caption{Hydrogen cyanide}
                \label{HCNfigure}
        \end{subfigure}
\\
        \begin{subfigure}[c]{0.2\textwidth}
                \includegraphics[width=2cm,height=2cm,keepaspectratio]{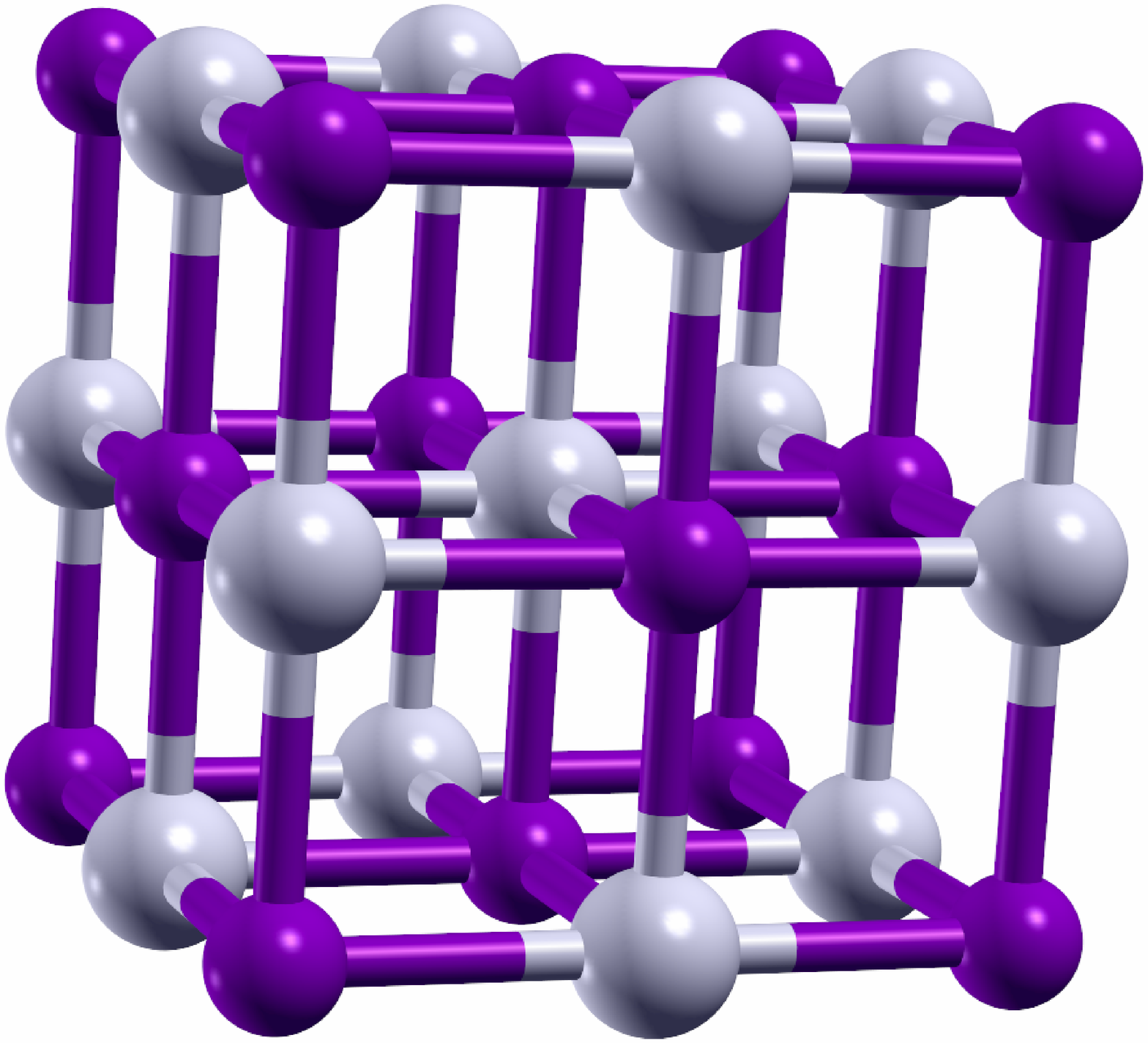}
                \caption{Lithium Hydride}
        \end{subfigure}
        \qquad
        \begin{subfigure}[c]{0.2\textwidth}
                \includegraphics[width=2cm,height=2cm,keepaspectratio]{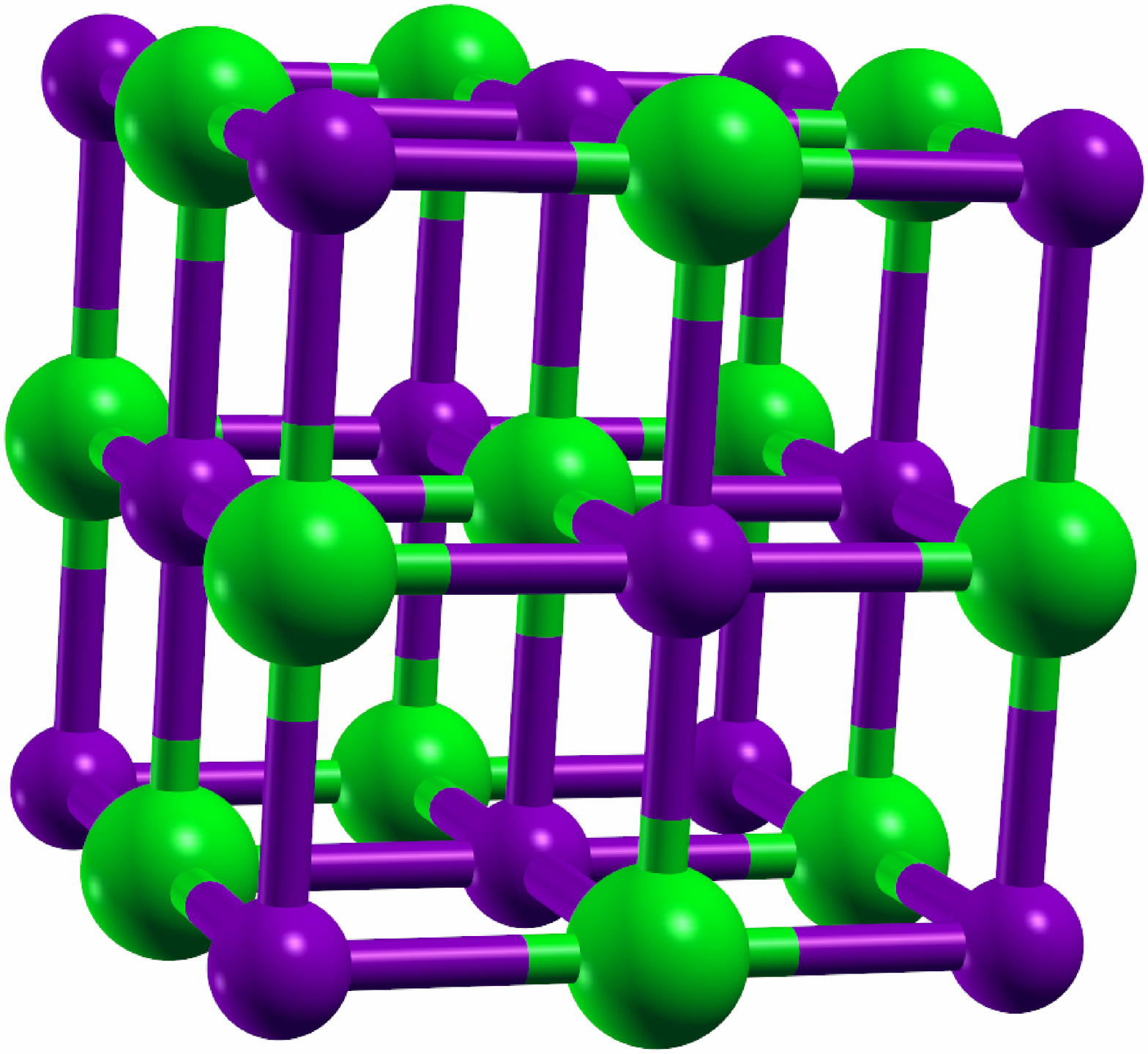}
                \caption{Lithium Fluoride}
        \end{subfigure}

        \begin{subfigure}[c]{0.2\textwidth}
                \includegraphics[width=2cm,height=2cm,keepaspectratio]{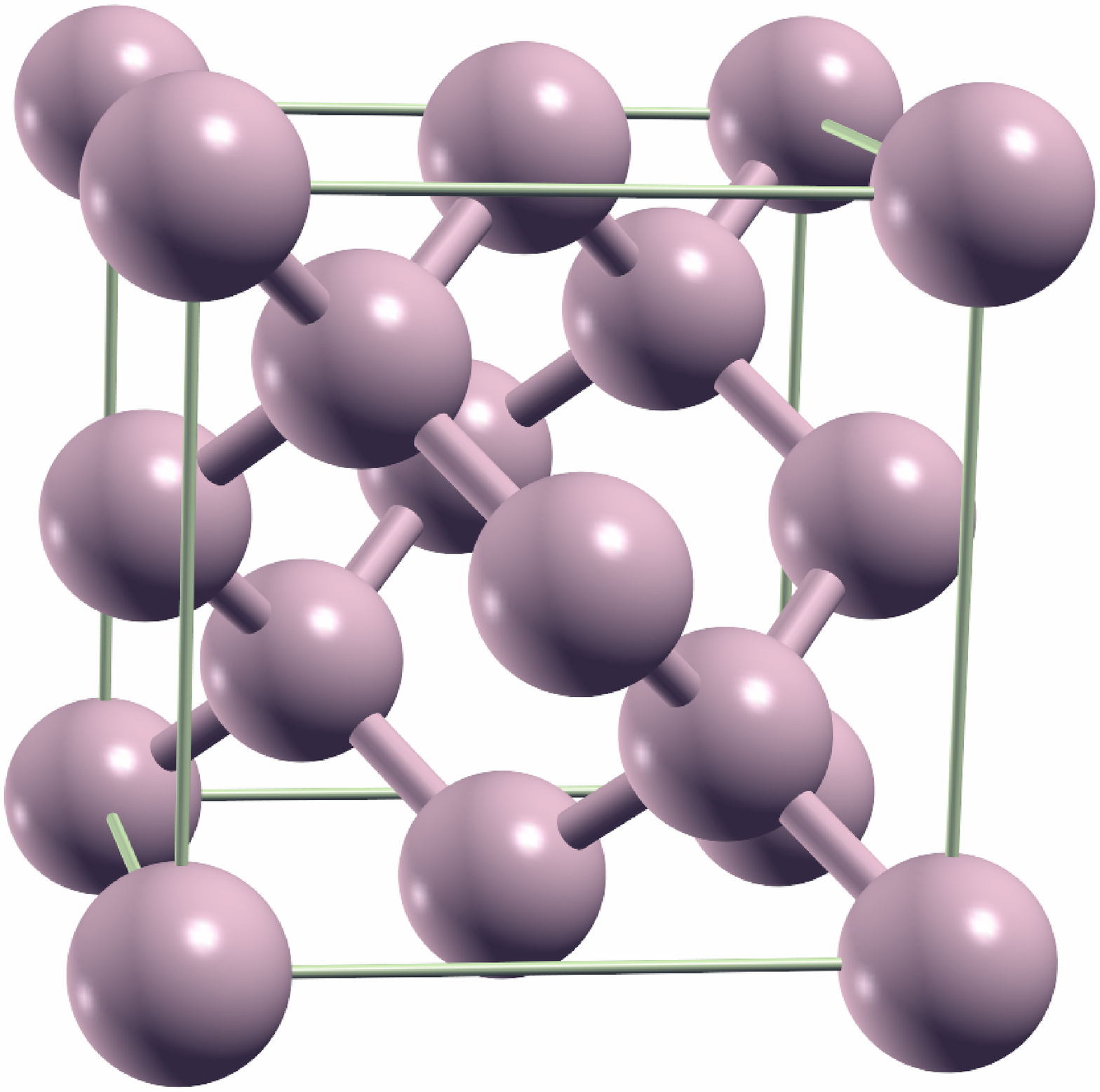}
                \caption{Silicon}
        \end{subfigure}
        \qquad
        \begin{subfigure}[c]{0.2\textwidth}
                \includegraphics[width=2cm,height=2cm,keepaspectratio]{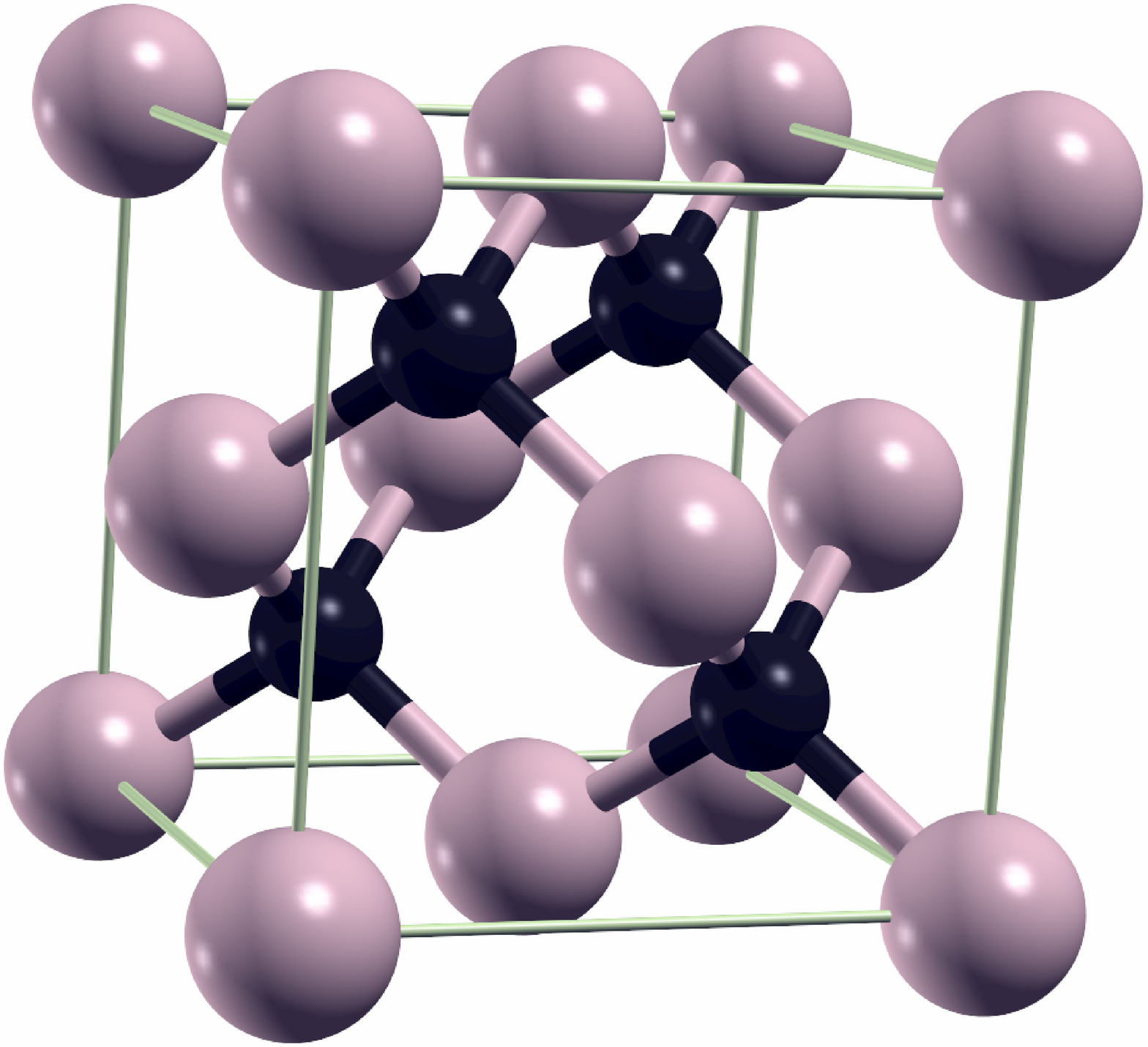}
                \caption{Silicon Carbide}
        \end{subfigure}
        \caption{Pictorial representation of unit cells of crystals used as a benchmark in this work.}
        \label{fig:unitcelles}
\end{figure}

\section{Computational details}
\label{sec:computational}
\subsection{Test systems}
\label{sec:systems}
A small but representative set of crystalline systems was chosen in order to cover the diverse types of chemical bond typical of solids (see Figure~\ref{fig:unitcelles}): rare-gas crystals (Ne, Ar), molecular solids (CO$_2$, HCN, NH$_3$), ionic crystals (LiH, LiF) and covalent semiconductors (Si, SiC). Metals were not considered in this study since the perturbative nature of the MP2 approach does not allow calculations on systems with a very small or zero band gap.\cite{PhysRevL_Mp2_metals}

The LiH crystal, being the simplest 3D crystal, has been the subject of much attention recently,\cite{Manby_LiH,MarGruPaiKre-JCP-09,Doll_LiH,Manby_LiH_New,PhysRevB.80.174114,PhysRevB.81.106101,DelHutVan-JCTC-12,usvyat2013} and convergence of the total MP2 energy with basis set was carefully investigated by some of us.\cite{LiH_MP2limit}
The three molecular crystals have been chosen {{in order to evaluate different nature of interactions which contribute to the total cohesion energy. This includes dispersive and electrostatic quadrupole-quadrupole interactions\cite{kihara} in CO$_2$ crystal and hydrogen bonds in NH$_3$ crystal.}}
{{The HCN crystal, instead, represents a simple example of a molecular crystal in which both dispersion and hydrogen bonding are present.
These particular systems have been studied by different authors with different approaches, including periodic post-Hartree-Fock methods in the last few years.\cite{MasUsvSchCiv-JCP-10,MasUsvCiv-CEC-10,MasCivUglGav-JPC-11,bygrave2012,DelHutVan-JCTC-12,muller2013,Hirata_CO2}}}
Rare-gas crystals are of general interest as purely dispersion-bonded
crystals, where the dominant role of electron correlation effects is well known.\cite{Halo:2009p30774,Halo:2009p30779,Casassa:2008p54747,PhysRevB.82.205111,Muller_Ar}

\begin{table*}
\caption{Structural information about the crystals used in this work. $n_{\text{ato}}$ is the total number of atoms per cell. 
{ In last column we report reference to the works each structure was taken from}.}
\begin{threeparttable}
\begin{tabular}{ccccccc} \hline\hline
                   & System   & $a;b;c$ (\AA) & Space group& $n_{\text{ato}}$ & Ref.\\\midrule
\hline
\rule{0pt}{3ex}
\multirow{2}{*}{Rare-gas}  & Ne  & 4.464         & Fm$\bar{3}$m          & 1                           & \cite{BatLosSim-PR-67,McC-JCP-74,EndShiSka-PRB-75} \\
                           & Ar  & 5.300         & Fm$\bar{3}$m          & 1                          & \cite{PetBatSim-PR-66,SchCraCheAzi-JCP-77} \\
                           \\
\rule{0pt}{3ex}
\multirow{3}{*}{Molecular} & $\text{CO}_2$      & 5.54& Pa$\bar{3}$           & 12                           & \cite{clarification-co2} \\
                           & $\text{NH}_3$ & 5.048 & $\text{P2}_1$3      & 16                    & \cite{KiePenBreClo-JCP-87} \\
                           & HCN  & 4.13; 4.85; 4.34& Imm2                  & 3                               & \cite{DulLip-AC-51} \\
                           \\
\rule{0pt}{3ex}
\multirow{2}{*}{Ionic}     & LiH & 4.084         & Fm$\bar{3}$m          & 2                       & \cite{Manby_LiH_New} \\
                           & LiF & 4.010         & Fm$\bar{3}$m          & 2                           & \cite{Lanboer} \\
                           \\
\rule{0pt}{3ex}
\multirow{2}{*}{Semiconductor } & Si & 5.430     & Fd$\bar{3}$m          & 2                         & \cite{TouKirTayDes} \\
                           & SiC    & 4.358     & F$\bar{4}$3m          & 2                       & \cite{TouKirTayLee} \\ \bottomrule
\hline\hline
\end{tabular}
\end{threeparttable}
\label{testsyst-info-tab}
\end{table*}

Details on the systems and the geometries we used in the calculations are reported in Table \ref{testsyst-info-tab}. The experimental lattice parameters, as indicated, were adopted in all cases.
For molecular crystals, internal coordinates were taken from Ref. \onlinecite{MasUsvSchCiv-JCP-10}, where they were obtained by a fixed-volume optimization of internal coordinates performed at the B3LYP-D* level.\cite{Civalleri:2008p37490}

\subsection{Parameters of the calculations}
All the periodic calculations were performed with development versions of the {\sc Crystal14}\cite{Dovesi2014} and {\sc Cryscor}\cite{PisBusCapCasDovMasZicSch-JCP-05,pisani2008,usvyat2010,PisSchCasUsvMasLorErb-PCCP-12} programs. Molecular calculations were performed either with the above codes or with {\sc Molpro}.\cite{MOLPRO-WIREs,MOLPRO_brief}
As for the {\sc Crystal} calculations, we adopted a homogeneous $8\times 8\times 8$ $\b{k}$-point sampling of the reciprocal space, and integral-screening tolerances set to $10^{-8}$,$10^{-8}$,$10^{-8}$,$10^{-20}$,$10^{-50}$. For the meaning of these thresholds we address the user to the {\sc Crystal14} user's manual;\cite{Cry-MAN-14} 
here we just point out that, as in our previous works,\cite{MasUsvSchCiv-JCP-10, LiH_MP2limit} we tightened the thresholds for the exchange integrals (last two numbers) with respect to defaults.

In {\sc Cryscor}, the fundamental input parameters refer to the locality ansatz. In particular the most relevant one is the selection of excitation domains assigned to each occupied orbital, { since in the present calculations the recently implemented orbital-specific-virtual method~\cite{osv} was not adopted}. 
Domains for rare-gas and ionic solids were chosen in order to consider PAOs
belonging to two coordinated shells of atoms around the occupied orbital, while for molecular crystals the domains coincide with the molecule to which the orbital belongs.
For semiconductor crystals, we selected the domains in order to take into account six tetrahedral units for a total amount of 26 atoms around each bond orbital.
The maximum pair distance considered for the evaluation of MP2 correlation energy was fixed at $12$ \AA.
Two-electron integrals within this range are evaluated efficiently in {\sc Cryscor} either by density fitting\cite{maschio2007,MasUsv-PRB-08,schutz2010} or multipolar expansion techniques -- if the inter-orbital distance exceeds 8 {\AA}.

\subsection{Basis sets} \label{bs-mol-crys}
As mentioned in the introduction, the {\sc Crystal} and {\sc Cryscor} programs use a basis set of Gaussian-type orbital functions centred on atoms to create atomic orbitals. 
In this study we mainly employed Dunning's cc-pVXZ.\cite{corr-cons-pol-bs}
Where possible, unmodified basis sets with X=D,T,Q, have been used. \\
Exceptions are:\cite{SuppMat}
\begin{itemize}
\item[-] LiH and LiF crystals: a suitably optimized cc-pVQZ basis set (See supplementary information for details) was adopted for Li in all calculations, regardless of the basis on H or F.
The basis sets indicated in the tables and figures refer, therefore, only to the latter atomic species, for which we used standard Dunning's basis sets.
\item[-] Si and SiC crystals: the cc-pVDZ basis for Si needed to be re-optimized in order to allow for SCF convergence. However, it was possible to use an unmodified cc-pVDZ basis on C.
\end{itemize}
Augmentation was made only for polarization shells, i.e. no diffuse s-type functions were included in any case. Hereafter, the prefix p-aug will be adopted. This augmentation scheme proved, in a number of cases, 
\cite{MasUsvSchCiv-JCP-10,MasUsvCiv-CEC-10,MasCivUglGav-JPC-11,Sharkas14} to be very effective in keeping the beneficial effects of polarization-augmented basis sets on the correlation energy while moderating the impact on the computational demands. 
 No dual basis-set scheme\cite{usvyat2010} is adopted in the present work, which would imply contributions from single excitations in the MP2 treatment\cite{Sharkas14}. As a consequence, the single-excitation contribution is always zero in the calculations presented here.
{ We did not attempt to extrapolate our results to the complete-basis-set limit (a recent work~\cite{RSH-extr} suggests an exponential convergence of the long-range correlation energy upon the cardinal number of the basis) since this would go beyond the scopes of the present work.}

\subsection{Cohesive energies}
\label{sec:cohenergy}

We computed the cohesive energy $E_{\text{coh}}(V)$ per $X$ (where $X$ can stand for either a molecule or an atom) at a given volume $V$ of the unit cell, as:
\begin{equation}
E_{\text{coh}}=\frac{E_{\text{bulk}}}{Z}-E_X^{[\text{gas}]},
\end{equation}
where $Z$ is either the number of molecular units or the number of atoms in the unit cell; $E_{\text{bulk}}$ and $E_X^{[\text{gas}]}$ are respectively the total energy per unit cell of the bulk system and the total energy of $X$ in the gas phase geometry -- which has been optimized at the B3LYP-D* level for molecular systems, for consistency with the adopted crystalline structures.
When $X$ refers to a single atom, as in the case of rare-gas crystals, obviously $E_X^{[\text{gas}]} = E_X^{[\text{bulk}]}$.

In order to correct for the basis set superposition error (BSSE) we adopted the standard Boys-Bernardi counterpoise (CP) method: \cite{BoyBer-MP-70}
\begin{equation}
E_{\text{coh}}^{\text{CP}}=E_{\text{coh}}+E_X^{[\text{bulk}]}-E_{X+\text{gh}}^{[\text{bulk}]}, \label{coh-ene-eq}
\end{equation}
where $E_X^{[\text{bulk}]}$ and $E_{X+\text{gh}}^{[\text{bulk}]}$ are the energies of $X$ in the crystalline bulk geometry without and with ghost functions, respectively.
Note that in the case of ionic crystals, cohesive energies are evaluated with
respect to the isolated atoms, not ions. The energies for the latter were
computed in the framework of the spin-unrestricted formalism, using the {\sc
  Molpro} code.\cite{MOLPRO-WIREs,MOLPRO_brief}

\begin{figure}
\begin{center}
\includegraphics[scale=0.55]{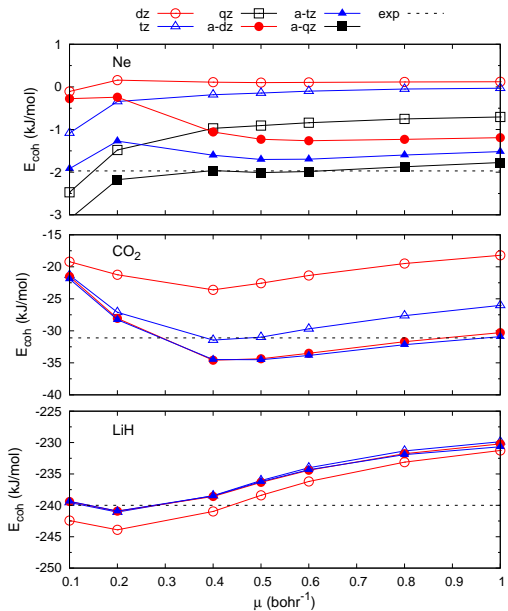}
\end{center}
\caption{Cohesive energy of the Ne, CO$_2$, and LiH crystals calculated with the RSHPBE+MP2 method as a function of the range-separation parameter $\mu$ for different basis sets.}
\label{fig:mudep}
\end{figure}

\begin{figure}
\begin{center}
\includegraphics[scale=0.25, angle=270]{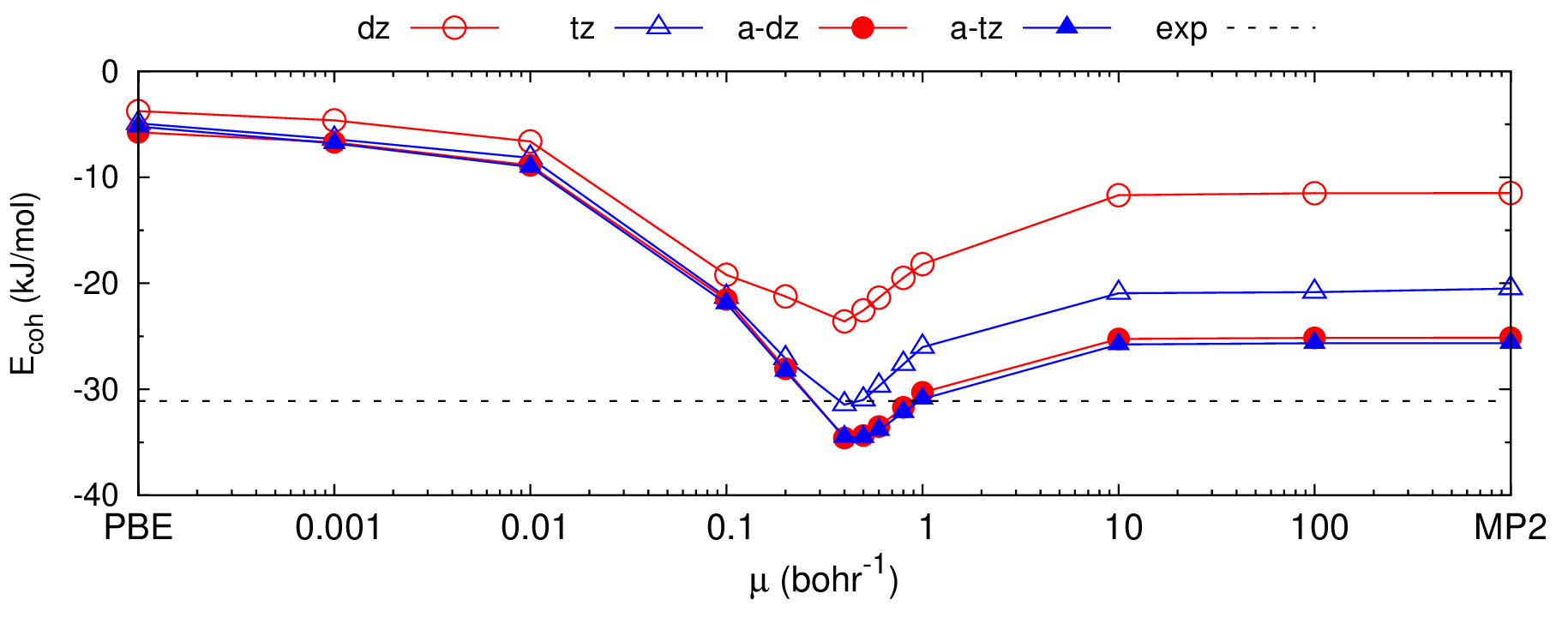}
\end{center}
\caption{Cohesive energy of the CO$_2$ crystal calculated with the RSHPBE+MP2 method as a function of the range-separation parameter $\mu$ for different basis sets.}
\label{fig:mudep-fullr}
\end{figure}

\begin{figure}
\begin{center}
\includegraphics[scale=0.55]{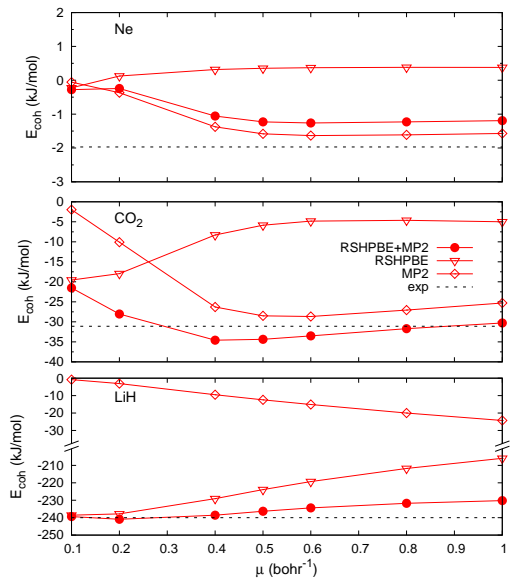}
\end{center}
\caption{Decomposition of the RSHPBE+MP2 cohesive energy of the Ne, CO$_2$, and LiH crystals in individual RSHPBE and MP2 contributions as a function of the range-separation parameter $\mu$ for the p-aug-cc-pVDZ basis set.}
\label{fig:decomp}
\end{figure}

\section{Dependence on the range-separation parameter}
\label{sec:mu}

The literature discussing the implementations of range-separated DFT approaches for molecular systems often adopts a value of $\mu=0.5$ bohr$^{-1}$ (if not otherwise specified, bohr$^{-1}$ will be dropped in the following sections) for the range-separation parameter (see Section \ref{sec:methods}). This originates from some benchmark on molecular systems, and is justified under the consideration that $\mu$ should correspond to the inverse of the average distance between valence electrons ({\it i.e. } twice the Seitz radius, giving around 1-2 bohr in valence regions).\cite{GerAng-CPL-05a}

Solids exhibit a wider variety of chemical bonds with respect to molecular
complexes: dispersion interactions, ionic, covalent, and metallic bond (the
latter we can not deal with, using the approaches presented in this work).
The crystalline environment is, by itself, different in nature from that of a molecular system, due to its infinite character and close packing of atoms.
For these reasons, we studied the dependence on the value of $\mu$. The range we considered was $0.1$--$1.0$ (corresponding to distances between electrons ranging from about 5 to 0.5 \AA) which represents a range of physically reasonable distances.

In Figure \ref{fig:mudep} the cohesive energy calculated with the RSHPBE+MP2 method is reported as a function of $\mu$ for three systems of our benchmark set: the Ne, CO$_2$, and LiH crystals. For each system, curves obtained with different basis sets are reported, as well as the experimental value.

In the case of the purely dispersion-bonded Ne crystal, the cohesive energy is
quite dependent on the basis set. It stems from the expected 
strong basis set dependence of the MP2 part, but also from the basis set dependence of the DFT part which 
is even stronger for this system. The
latter can be probably explained by the very small
scale of the interaction energy (of an order of just a kJ/mol) and 
{ a delicate balance between dispersion attraction and exchange repulsion }
which needs to be captured in a proper way.  At the same time, for
any given basis set, the energy is almost independent of $\mu$ above
$\mu \gtrsim0.5$.
In particular, with the largest basis set (p-aug-cc-pVQZ), RSHPBE+MP2
gives an accurate cohesive energy for $\mu \gtrsim 0.4$.

For the other two systems, the basis set dependence 
is virtually negligible at the scale of the
interaction energy, provided
the polarization-augmented basis sets are used.
For the CO$_2$ crystal (middle panel of Figure \ref{fig:mudep}), the curves of the cohesive energy exhibit a minimum at $\mu \approx0.4-0.5$ for all basis sets.
 Near this minimum, RSHPBE+MP2 slightly overbinds with the augmented basis sets.
 The obtained cohesive energy is much more accurate than pure PBE ($\mu=0$), as shown in Figure \ref{fig:mudep-fullr} where we reported the same curves in an extended range of $\mu$. The effect of the diffuse basis functions on the RSHPBE+MP2 cohesive energy is important for $\mu \gtrsim 0.4$ and seems converged with the p-aug-cc-pVDZ basis set. 
{
For $\mu \lesssim 0.2$ the role of many-body correlation effects, not included in the exchange-correlation functional, becomes insignificant and the impact of augmentation of the basis set becomes very small.
}
For the LiH crystal, the cohesive energy curves are quite flat between $\mu=0.1$ and $0.5$, with a minimum at around $\mu \approx 0.2$. Near the minimum, RSHPBE+MP2 gives an accurate cohesive energy with basis sets larger than the cc-pVDZ basis set.

In Figure \ref{fig:decomp} the decomposition of the RSHPBE+MP2/p-aug-cc-pVDZ cohesive energy curves of Figure \ref{fig:mudep} in individual RSHPBE and MP2 contributions is reported. It is interesting to observe how the absolute value of the MP2 contribution increases for Ne and CO$_2$ when increasing $\mu$ from 0.1 to about 0.5, and then saturates. For LiH, the absolute value of the MP2 contribution increases almost linearly as a function of $\mu$. 
This can be rationalized in the following way. By increasing the $\mu$
parameter, one progressively includes more short-range MP2 correlation. For
small values of $\mu$ it still means adding more dispersion, leading to
progressive growth of the magnitude of the attractive MP2
contribution. However, at some point, $\mu$ becomes so large, that the
short-range intra-molecular MP2 component of the interaction energy, which is usually repulsive in
molecular crystals,\cite{muller2013} starts
contributing. Further on it even outruns the short-range dispersion, whose
accumulation slows down due to the packing effects,
and the MP2 contribution curve for Ne and CO$_2$ turns into a slightly decaying regime.
In contrast to the molecular crystals, for LiH the short-range correlation is
stabilizing (correlation substantially strengthens binding even in LiH
molecule\cite{Manby_LiH_New}), so in this system the magnitude of the MP2
contribution always grows with increase of the $\mu$ parameter.

From the present results, we conclude that the value of $\mu=0.5$ is reasonable for the solids considered. Hence, as a conservative choice based on previous experience on molecular systems, we chose this value of $\mu$ for the wider benchmark of the range-separated double hybrids carried out in Section \ref{sec:results}. Notably, for that value of $\mu$, range-separated double hybrids show a similar basis set dependence as full MP2 for the basis sets considered here.

\section{A wider benchmark of range-separated double hybrids in solids}
\label{sec:results}

\begin{table*}
\footnotesize
\begin{threeparttable}
\caption{Benchmark of range-separated double-hybrid methods on cohesive energies ($E_\text{coh}$ in kJ/mol) of crystalline systems using double-zeta quality basis sets. The RSHLDA+MP2, RSHPBE+MP2, RSHLDA+SCS, and RSHPBE+SCS results are obtained with a value of the range-separation parameter of $\mu=0.5$. The statistical indicators calculated are: mean errors (MEs), error variances ($\sigma^2$), mean absolute error (MAEs), and mean absolute relative errors (MAREs).} \label{bench-dz}
\begin{tabular}{ccccccccccccccccccccccccccc}
\hline\hline

\multirow{2}{*}{Crystal} & \multicolumn{2}{c}{LDA} && \multicolumn{2}{c}{PBE} && \multicolumn{2}{c}{RSHLDA+MP2} &&  \multicolumn{2}{c}{RSHPBE+MP2} && \multicolumn{2}{c}{RSHLDA+SCS} &&  \multicolumn{2}{c}{RSHPBE+SCS} && \multicolumn{2}{c}{MP2} && \multicolumn{2}{c}{SCS} & \multirow{2}{*}{Exp.} & \multirow{2}{*}{Ref.$^{a)}$} \\
\cline{2-3}\cline{5-6}\cline{8-9}\cline{11-12}\cline{14-15}\cline{17-18}\cline{20-21}\cline{23-24}
& $E_\text{coh}$ & \% && $E_\text{coh}$ & \% && $E_\text{coh}$ & \% && $E_\text{coh}$ & \% && $E_\text{coh}$ & \% && $E_\text{coh}$ & \% && $E_\text{coh}$ & \% && $E_\text{coh}$ & \% & & \\ \midrule

                        Ne    &-0.31& 84&&-0.27&86 &&+0.11&106&&+0.10&105&&+0.18&109&&+0.18&109&&+0.22&111&&+0.28&114&-1.97
 &\cite{McC-JCP-74} \\
                        Ar    &-6.84&12 &&+2.36&131&&+1.10&114&&+0.87&111&&+2.00&126&&+1.77&123&&+3.13&140&&+4.07&153&-7.73
 &\cite{SchCraCheAzi-JCP-77} \\

\rule{0pt}{3ex} $\text{CO}_2$ &-27.8&11 &&-3.8 &88 &&-21.9&30 &&-22.6&27 &&-17.6&43 &&-18.2& 41&&-11.5& 63&& -7.9& 75&-31.1
 &\cite{ChiAcr-JPCRD-02} \\
$\text{NH}_3$                 &-55.2&52 &&-26.9&26 &&-32.1&12 &&-32.8&10 &&-29.1&20 &&-29.7& 18&&-24.2&33 &&-19.5& 46&-36.3
 &\cite{ChiAcr-JPCRD-02} \\
HCN                           &-51.4&21 &&-28.0&34 &&-40.1&6  &&-40.4&5  &&-36.3&15 &&-36.6& 14&&-31.7& 26&&-27.2& 36&-42.6
 &\cite{ChiAcr-JPCRD-02} \\

        \rule{0pt}{3ex} LiH   &-265 &10 &&-236 & 2 &&-245 &2  &&-238 & 1 &&-244 & 2 &&-237 & 1 &&-226 & 6 &&-229 & 5 &-240
&\cite{NJT} \\
        LiF                   &-482 &12 &&-425 &1  &&-454 &6  &&-438 &2  &&-452 & 5 &&-438 &2  &&-432 &0  &&-427 & 1 &-430
&\cite{NJT} \\

        \rule{0pt}{3ex} Si    &-506 &12 &&-433 & 4 &&-493 & 9 &&-490 & 8 &&-469 & 4 &&-465 &3  &&-433 & 4 &&-404 &11 &-452
&\cite{NJT} \\
        SiC                   &-697 &12 &&-603 & 4 &&-648 & 4 &&-638 &  2&&-633 &0  &&-623 &0  &&-585 & 6 &&-559 &11 &-625
&\cite{NJT} \\
\rule{0pt}{3ex} $\sigma^2$    &12.1 &   &&5.0  &   &&6.0  &   &&4.8  &   &&3.9  &   &&2.7  &   &&6.0  &   &&9.9  &   &    & \\
ME                            &-25.0&   &&12.6 &   &&-7.4 &   &&-3.6 &   &&-1.3 &   &&2.4  &   &&14.1 &   &&21.9 &   &    & \\
MAE                           &26.3 &   &&12.6 &   &&13.3 &   &&9.5  &   &&10.0 &   &&7.0  &   &&14.5 &   &&21.9 &   &    & \\
MARE                          &     &25 &&     &42 &&     &32 &&     &30 &&     &36 &&     &35 &&     &43 &&     &50 &    & \\

\hline\hline
\end{tabular}
\begin{tablenotes}
\item{$^{a)}$ The experimental sublimation energies of molecular crystals taken from Ref. \onlinecite{ChiAcr-JPCRD-02} were corrected for zero-point energy (ZPE) and thermal effects at $298$ K by a constant $2RT$ contribution.\cite{Gav-MSMSE-02}
Atomization energies of semiconductors and ionic crystals were corrected for ZPE in accordance to the zero-point anharmonic expansion correction (values from Ref.~\onlinecite{SchHarKre-JCP-11}) }\\
\end{tablenotes}
\end{threeparttable}
\end{table*}

\begin{table*}
\footnotesize
\begin{threeparttable}
\caption{Same as Table \ref{bench-dz} with polarization-augmented double-zeta quality basis sets.} \label{bench-adz}
\begin{tabular}{ccccccccccccccccccccccccccc}
\hline\hline

\multirow{2}{*}{Crystal} & \multicolumn{2}{c}{LDA} && \multicolumn{2}{c}{PBE} && \multicolumn{2}{c}{RSHLDA+MP2} &&  \multicolumn{2}{c}{RSHPBE+MP2} && \multicolumn{2}{c}{RSHLDA+SCS} &&  \multicolumn{2}{c}{RSHPBE+SCS} && \multicolumn{2}{c}{MP2} && \multicolumn{2}{c}{SCS} & \multirow{2}{*}{Exp.} & \multirow{2}{*}{Ref.$^{a)}$} \\
\cline{2-3}\cline{5-6}\cline{8-9}\cline{11-12}\cline{14-15}\cline{17-18}\cline{20-21}\cline{23-24}
& $E_\text{coh}$ & \% && $E_\text{coh}$ & \% && $E_\text{coh}$ & \% && $E_\text{coh}$ & \% && $E_\text{coh}$ & \% && $E_\text{coh}$ & \% && $E_\text{coh}$ & \% && $E_\text{coh}$ & \% & & \\ \midrule

Ne                            &-0.44& 78&&-0.40&86 &&-1.24&37 &&-1.23&38 &&-0.87&56 &&-0.86&56 &&-1.10&44 &&-0.74&62 &-1.97
&\cite{McC-JCP-74} \\
Ar                            &-11.9&54 &&+0.42&105&&-7.50&3  &&-7.65&  1&&-4.88&37 &&-5.03&35 &&-6.45&17 &&-3.57&54 &-7.73
&\cite{SchCraCheAzi-JCP-77} \\

\rule{0pt}{3ex} $\text{CO}_2$ &-32.9&6  &&-5.8 &81 &&-33.6&8  &&-34.4&11 &&-17.6&43 &&-27.5& 12&&-25.2& 19&&-18.5& 41&-31.1
 &\cite{ChiAcr-JPCRD-02} \\
$\text{NH}_3$                 &-55.6&53 &&-26.2&28 &&-38.8&7  &&-39.7&9  &&-29.1&20 &&-35.0& 4 &&-31.8&12 &&-25.4& 30&-36.3
 &\cite{ChiAcr-JPCRD-02} \\
HCN                           &-54.9&29 &&-29.7&30 &&-48.4&14 &&-48.7&14 &&-36.3&15 &&-43.4& 2 &&-41.4& 3 &&-35.0& 18&-42.6
 &\cite{ChiAcr-JPCRD-02} \\

\rule{0pt}{3ex} LiH           &-263 &10 &&-233 & 3 &&-243 &1  &&-236 & 2 &&-241 & 0 &&-235 & 2 &&-225 & 6 &&-229 & 5 &-240
&\cite{NJT} \\
LiF                           &-482 &12 &&-426 &1  &&-454 &6  &&-440 &2  &&-453 & 5 &&-440 &2  &&-435 &1  &&-429 & 0 &-430
&\cite{NJT} \\

\rule{0pt}{3ex} Si            &-509 &13 &&-436 &4  &&-474 & 5 &&-474 & 5 &&-453 &0  &&-448 &1  &&-425 &6  &&-395 &13 &-452
&\cite{NJT} \\
SiC                           &-701 &12 &&-607 &3  &&-652 &4  &&-642 & 3 &&-638 &2  &&-627 &0  &&-599 & 4 &&-577 &8  &-625
&\cite{NJT} \\
\rule{0pt}{3ex} $\sigma^2$    &12.6 &   &&4.5  &   &&4.8  &   &&3.4  &   &&3.0  &   &&1.4  &   &&4.6  &   &&8.6  &   &     & \\
ME                            &-27.1&   &&11.4 &   &&-9.5 &   &&-6.3 &   &&-3.1 &   &&0.5  &   &&8.5  &   &&17.1 &   &     & \\
MAE                           &27.5 &   &&11.4 &   && 9.8 &   &&7.4  &   &&5.4  &   &&3.4  &   &&9.6  &   &&17.1 &   &     & \\
MARE                          &     &30 &&     &37 &&     &9  &&     &9  &&     &13 &&     &13 &&     &12 &&     &25 &     & \\

\hline\hline
\end{tabular}
\begin{tablenotes}
\item{$^{a)}$ See footnote in Table \ref{bench-dz}.}\\
\end{tablenotes}
\end{threeparttable}
\end{table*}

The results reported in Section \ref{sec:mu} give a first hint that range-separated double hybrids might provide cohesive energy of solids that are more accurate than the ones obtained by pure DFT or pure MP2, as generally observed for molecular systems.
 In this Section,  we further benchmarked the range-separated double hybrids, with double-zeta quality basis sets, on the set of solids described in Section \ref{sec:systems}.
In Table \ref{bench-dz} cohesive energies are compiled for the cc-pVDZ basis set and the following methods:
\begin{itemize}
\item[-] Pure DFT (LDA, PBE), corresponding to $\mu=0$;
\item[-] Pure MP2 and SCS-MP2, corresponding to $\mu=\infty$;
\item[-] Range-separated combinations of the above with $\mu=0.5$.
\end{itemize}
For each system the zero-point-energy corrected experimental cohesive energy is also provided, and a number of statistical indicators are calculated.

Looking at the mean errors (MEs), the mean absolute errors (MAEs), or the error variances ($\sigma^2$), it is seen that the range-separated double hybrids globally perform better for cohesive energies than the respective pure DFT and pure MP2 methods. The mean absolute relative errors (MAREs) also confirm this trend, with the exception of pure LDA which gives better cohesive energies for Ar and CO$_2$ with the cc-pVDZ basis set compared to the other methods, leading to the smallest MARE despite being less accurate than the other methods for ionic and semiconducting crystals. As a matter of fact, the large difference in behavior on individual systems observed between pure LDA and PBE is greatly smoothened in the range-separated double hybrids which yield similar results for the two density functionals. Yet, on average, the PBE-based range-separated double hybrids gives slightly more accurate cohesive energies than the LDA-based ones. The SCS variants of the range-separated double hybrids perform better that the non-SCS ones for the semiconducting crystals, but worse for the rare-gas and molecular crystals.

Results for the p-aug-cc-pVDZ basis are analogously reported in Table \ref{bench-adz}. Similar trends are observed even though the results are more mixed. In comparison to pure PBE and pure MP2, the RSHPBE+MP2 method provides cohesive energies that are more accurate for Ne, Ar, CO$_2$, NH$_3$, and about as accurate for LiH, LiF, Si and SiC. 
The HCN crystal is a challenging system: it is known that pure MP2 (with large basis sets) tends to overbind this crystal,\cite{muller2013} and the RSHLDA+MP2 or RSHPBE+MP2 method seems to accentuate this behavior.

Concerning molecular crystals, results can be compared with density-scaled double hybrids reported in Ref. \onlinecite{Sharkas14}. 
For HCN, the comparison is more straightforward since the same basis set was used in both cases.
With the DS1DH-PBEsol ($\lambda = 0.80$) functional, the computed cohesive energies are -33.6 
and -39.3 kJ/mol for the cc-pVDZ and p-aug-cc-pVDZ basis sets, respectively, remarkably smaller than
RSHLDA+MP2 and RSHPBE+MP2 methods. A similar behaviour is observed for CO$_2$ and NH$_3$, even if
the basis sets are not exactly the same. For DS1DH-PBE, the cohesive energies are even smaller.

\section{Conclusions}
\label{sec:conclusions}

\begin{figure}
\begin{center}
\includegraphics[scale=0.7]{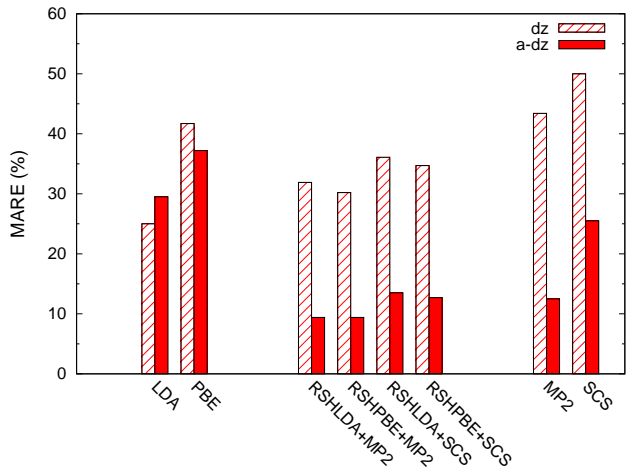}
\end{center}
\caption{Summary of the performance of methods tested in this work on crystalline system, measured by their MAREs (cf. Tables \ref{bench-dz} and \ref{bench-adz}). Empty lined bars refer to the cc-pVDZ basis set and fully colored bars refer to the p-aug-cc-pVDZ basis set.}
\label{fig:summary-mare}
\end{figure}

We implemented and tested range-separated double-hybrids methods in the {\sc Crystal} and {\sc Cryscor} programs for the study of crystalline solids. 
The approaches considered include either LDA- or PBE-type density functionals for short-range electron-electron interactions, and a local MP2 correlation correction for the long-range electron-electron interactions either in its standard or its SCS form.

The value of $\mu=0.5$  bohr$^{-1}$ for the range-separation parameter commonly adopted for molecular systems was found to be also a reasonable choice for solids and was thus adopted for this study.
The range-separated double-hybrids methods have been tested on a significant test set of cohesive energies of nine prototypical crystalline systems.
A summary of the results is provided in Figure \ref{fig:summary-mare}. 
With double-zeta correlation consistent basis sets, either augmented with diffuse polarization functions or not, the range-separated double hybrids are globally more accurate than the respective pure DFT or MP2 calculations. As for pure MP2, the effect of augmentation of the basis set with diffuse  polarization functions is important for the range-separated double hybrids, reducing the MARE values by about a factor of three with respect to non-augmented results. Overall, the SCS variants of the range-separated double-hybrids appear to be less accurate than the non-SCS ones, but the reader shall be aware that the present benchmark set does not include stacking-type interactions, where SCS is expected to bring more significant improvements.

Future developments of range-separated DFT approaches for solids might include
the implementation of other flavors of the methods (such as different
short-range density functionals~\cite{GolErnMoeSto-JCP-09} or different
long-range electron correlation
methods~\cite{TouZhuSavJanAng-JCP-11,AngLiuTouJan-JCTC-11}). In particular,
the notorious overestimation of dispersion by MP2 in highly polarizable
systems, which can affect, as observed in this work, also the range-separated double-hybrids employing MP2
as the long-range model, can be cured by substituting the
latter with approximate coupled-cluster models, containing only the low-order slowly decaying terms.\cite{masur2013,schutz2014,werner2015}

We plan also to gain more experience on calculations tackling ``real-life'' problems.
The performance on diverse properties other than cohesive energies will be of interest. One notable example is the relative stability of crystalline polymorphs,
\cite{Presti_2014,GeF2,halo2011b} which traditionally represent a tough
challenge for quantum chemistry methods. Another quantity of interest, where
the range-separated double hybrids can become particularly effective, is physisorption on surfaces or in
porous crystals. Indeed, since the short-range part of the
interaction, including the intra-host and intra-adsorbate components, is in this scheme
described by DFT, the MP2 treatment can be reduced to the inter-host-adsorbate
pairs only,\cite{Reinhardt2012} making such calculations computationally very efficient.

\section*{Acknowledgments}
We are grateful to Andreas Savin, Roberto Dovesi and Martin Sch\"{u}tz for fruitful discussions.
D. U. gratefully acknowledges financial support by Deutsche Forschungsgemeinschaft (grants US-103/1-1, US-103/1-2).

\begin{figure}
\begin{center}
\includegraphics[scale=0.21]{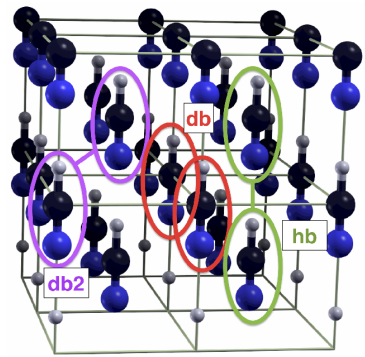}
\end{center}
\caption{The HCN dimers extracted from the bulk structure.}
\label{fig:hcnbulk}
\end{figure}

\appendix
\section{Molecular dimers cut from the bulk}
\label{appendix:dimers}
In order to highlight the peculiarities of the crystalline case with respect to the molecular one, we have performed the same kind of calculations as presented in Tables \ref{bench-dz} and \ref{bench-adz} on molecular dimers. Since our aim is not to discuss the performance of range-separated double hybrids on molecular complexes, which has been widely explored in the literature, we have extracted dimers from the bulk structures without re--optimizing the geometries. This has been done for rare-gas, ionic, and molecular crystals.
As a reference, we have performed calculations at the same geometry with coupled cluster singles, doubles, and perturbative triples (CCSD(T)) using an p-aug-cc-pV5Z basis set.

\begingroup
\squeezetable
\begin{table*}
\caption{Molecular complexes: benchmark of range-separated double-hybrid methods on interaction energies ($E_\text{int}$ in kJ/mol) of dimers cut out from bulk crystals using the cc-pVDZ basis set. The RSHLDA+MP2, RSHPBE+MP2, RSHLDA+SCS, and RSHPBE+SCS results are for a value of the range-separation parameter of $\mu=0.5$. Reference CCSD(T) calculations are for an aug-cc-pV5Z basis set. ``hb'', ``db'' and ``db2'' label three different HCN dimers, cfr. Figure \ref{fig:hcnbulk}}
 \label{bench-dim-dz}
\begin{tabular}{cccccccccccccccccccccccccc}
\hline\hline

\multirow{2}{*}{System} & \multicolumn{2}{c}{LDA} && \multicolumn{2}{c}{PBE} && \multicolumn{2}{c}{RSHLDA+MP2} &&  \multicolumn{2}{c}{RSHPBE+MP2} && \multicolumn{2}{c}{RSHLDA+SCS} &&  \multicolumn{2}{c}{RSHPBE+SCS} && \multicolumn{2}{c}{MP2} && \multicolumn{2}{c}{SCS} & \multirow{2}{*}{CCSD(T)} \\ \cline{2-3}\cline{5-6}\cline{8-9}\cline{11-12}\cline{14-15}\cline{17-18}\cline{20-21}\cline{23-24}
& $E_\text{int}$ & \% && $E_\text{int}$ & \% && $E_\text{int}$ & \% && $E_\text{int}$ & \% && $E_\text{int}$ & \% && $E_\text{int}$ & \% && $E_\text{int}$ & \% && $E_\text{int}$ & \% & \\ \hline

$\left[\text{Ne}\right]_2$                     &-0.027&85 &&-0.024&86 &&+0.013&107&&+0.012&107&&+0.018&110&&+0.018&110&&+0.022&112&&+0.026&115&-0.178\\
$\left[\text{Ar}\right]_2$                   &-0.64 &19 &&+0.06 &112&&+0.15 &128&&+0.13 &124&&+0.21 &139&&+0.19 &136&&+0.32 &159&&+0.38 &171&-0.54 \\

\rule{0pt}{3ex}$\left[\text{CO}_2\right]_2$  &-4.7 &116&&-0.9 &59 &&-2.9 &30 &&-3.0 &35 &&-2.2 &2 &&-2.4 &7  &&-1.4  &34 &&-0.9  &59 &-2.2  \\
$\left[\text{NH}_3\right]_2$          &-18.3 &219&&-10.7 &87 &&-9.9  &73 &&-10.0 &74 &&-9.4  &63 &&-9.4  &64 &&-7.8  &36 &&-6.8  &18 &-5.7  \\
$\left[\text{HCN}\right]_2$ / hb                &-27.0 &173&&-17.5 &76 &&-20.4 &106&&-20.5 &107&&-19.5 &97 &&-19.6 &98 &&-16.3 &64 &&-14.6 &48 &-9.9  \\
$\left[\text{HCN}\right]_2$ / db                &+3.8  &74 &&+4.3  &95 &&+4.5  &105&&+4.4  &101&&+4.8  &119&&+4.7  &115&&+4.3  &96 &&+4.5  &107&+2.2  \\
$\left[\text{HCN}\right]_2$ / db2               &-0.9  &321&&+2.1  &418&&+1.8  &349&&+1.7  &310&&+2.4  &484&&+2.2  &446&&+2.3  &467&&+2.8  &608&+0.4  \\

\rule{0pt}{3ex} Li--H   &-715  &11 &&-712  &11 &&-702  &10 &&-703  &10 &&-702  &10 &&-703  &10 &&-697  &9  &&-698  &9  &-641  \\
Li--F                   &-767  &13 &&-762  &13 &&-745  &10 &&-745  &10 &&-745  &10 &&-745  &10 &&-737  &9  &&-737  &9  &-676  \\

\rule{0pt}{3ex} $\sigma^2$ &13.2&&&12.4  &   &&10.3  &   &&10.4  &   &&10.3  &   &&10.4  &   &&9.3   &   &&9.3   &   &      \\
ME                    &-21.8 &   &&-18.5 &   &&-15.8 &   &&-16.0 &   &&-15.5 &   &&-15.7 &   &&-13.7 &   &&-13.3 &   &      \\
MAE                   &22.2  &   &&19.4  &   &&16.6  &   &&16.7  &   &&16.5  &   &&16.6  &   &&14.6  &   &&14.6  &   &      \\
MARE                  &      &115&&      &106&&      &102&&      &98 &&      &115&&      &111&&      &110&&      &127&      \\

\hline\hline
\end{tabular}
\end{table*}
\endgroup

\begingroup
\squeezetable
\begin{table*}
\caption{Molecular complexes: same as Table \ref{bench-dim-dz} with the p-aug-cc-pVDZ basis set.} 
\label{bench-dim-adz}
\begin{tabular}{cccccccccccccccccccccccccc}
\hline\hline

\multirow{2}{*}{System} & \multicolumn{2}{c}{LDA} && \multicolumn{2}{c}{PBE} && \multicolumn{2}{c}{RSHLDA+MP2} &&  \multicolumn{2}{c}{RSHPBE+MP2} && \multicolumn{2}{c}{RSHLDA+SCS} &&  \multicolumn{2}{c}{RSHPBE+SCS} && \multicolumn{2}{c}{MP2} && \multicolumn{2}{c}{SCS} & \multirow{2}{*}{CCSD(T)}\\ \cline{2-3}\cline{5-6}\cline{8-9}\cline{11-12}\cline{14-15}\cline{17-18}\cline{20-21}\cline{23-24}
& $E_\text{int}$ & \% && $E_\text{int}$ & \% && $E_\text{int}$ & \% && $E_\text{int}$ & \% && $E_\text{int}$ & \% && $E_\text{int}$ & \% && $E_\text{int}$ & \% && $E_\text{int}$ & \% & \\ \hline

$\left[\text{Ne}\right]_2$      &-0.038&79 &&-0.035&80 &&-0.080&55 &&-0.081&55 &&-0.055&69 &&-0.055&69 &&-0.072&59 &&-0.047&74 &-0.178\\
$\left[\text{Ar}\right]_2$      &-1.07 &98 &&-0.14 &74 &&-0.34 &37 &&-0.37 &32 &&-0.17 &69 &&-0.19 &64 &&-0.22 &59 &&-0.03 &95 &-0.54 \\

\rule{0pt}{3ex}$\left[\text{CO}_2\right]_2$ &-5.8 &164&&-1.4 &38 &&-4.6 &110&&-4.8 &116&&-3.6 &66 &&-3.8 &72 &&-3.4 &56 &&-2.4  &11 &-2.2  \\
$\left[\text{NH}_3\right]_2$         &-17.9 &212&&-10.2 &78 &&-11.1 &94 &&-11.2 &95 &&-10.3 &79 &&-10.3 &80 &&-9.4  &64 &&-8.0  &40 &-5.7  \\
$\left[\text{HCN}\right]_2$ / hb                &-27.9 &182&&-18.1 &83 &&-22.1 &123&&-22.1 &123&&-20.9 &111&&-21.0 &112&&-18.1 &82 &&-16.2 &63 &-9.9  \\
$\left[\text{HCN}\right]_2$ / db                &+3.7  &70 &&+4.3  &99 &&+4.3  &98 &&+4.2  &94 &&+4.8  &118&&+4.7  &114&&+3.7  &68 &&+4.1  &88 &+2.2  \\
$\left[\text{HCN}\right]_2$ / db2               &-1.5  &468&&+1.9  &362&&+0.7  &66 &&+0.5  &19 &&+1.5  &260&&+1.3  &215&&+0.8  &100&&+1.7  &316&+0.4  \\

\rule{0pt}{3ex} Li--H   &-720  &12 &&-717  &12 &&-707  &10 &&-708  &10 &&-707  &10 &&-708  &10 &&-700  &9  &&-700  &9  &-641  \\
Li--F                   &-765  &13 &&-760  &12 &&-744  &10 &&-745  &10 &&-744  &10 &&-744  &10 &&-736  &9  &&-736  &9  &-676  \\

\rule{0pt}{3ex} $\sigma^2$ &13.4& &&12.6 &   &&10.6  &   &&10.7  &   &&10.6  &   &&10.7  &   &&9.4   &   &&9.4   &   &      \\
ME                    &-22.5 &   &&-19.1 &   &&-16.9 &   &&-17.1 &   &&-16.5 &   &&-16.7 &   &&-14.7 &   &&-14.1 &   &      \\
MAE                   &22.9  &   &&19.8  &   &&17.4  &   &&17.5  &   &&17.2  &   &&17.3  &   &&14.9  &   &&14.6  &   &      \\
MARE                  &      &144&&      &93 &&      &67 &&      &62 &&      &88 &&      &83 &&      &56 &&      &78 &      \\

\hline\hline
\end{tabular}
\end{table*}

Tables \ref{bench-dim-dz} and \ref{bench-dim-adz} compile the results for the cc-pVDZ and p-aug-cc-pVDZ basis sets.
In the case of HCN three different types of dimer were chosen, to highlight the different types of interaction present in the cluster. 
In Figure \ref{fig:hcnbulk} a portion of the bulk crystal is reported, where some monomers are labeled: one purely hydrogen-bonded (``hb''), and two dispersion-bonded ( ``db'' and ``db2'' ).
In the case of LiH and LiF the cohesive energy is computed with respect to the neutral atoms.

Although the test set is not the same as in Section \ref{sec:results}, and is strongly biased towards molecular and dispersion-bonded systems, some comparisons on MAEs and MAREs can be made. At a first glance it clearly appears that these two indicators are significantly higher than in the bulk systems. 

For the rare-gas dimers, the results are tremendously improved by the addition of diffuse functions in the basis set. This effect is almost exactly parallel to what was observed for bulk crystals, and the same can be said about the performance of different methods for these systems.

The molecular dimers show a completely different picture: here relative errors are quite large in all cases, so that these errors dominate the overall statistics. 
It is indeed clear, from inspection of different dimers from the HCN crystal structure, how good (excellent, in the case of pure LDA) results for the bulk come from error cancellations -- overestimation of hydrogen bonds and underestimation of dispersion interactions.
One must not be surprised to find positive value for the interaction between some dimers, because they are not at the equilibrium geometry.

Looking at ionic systems -- LiH and LiF -- we see that the results seem almost insensitive to the method and to the augmentation of the basis set. We attribute this behavior to the strong, extremely short-ranged character of the ionic interaction. At the same time, recalling the results from Section \ref{sec:results}, long-range electron correlation effects (dispersion) are key to a correct description of the bulk.



\begin{thebibliography}{139}
\bibitem[{cla()}]{clarification-co2}
\bibinfo{note}{The experimental value extrapolated to 0 K was originally
  reported to be 5.54 (Ref. \cite{KeeKoh-Phy-34}). It has been later pointed
  out (Ref. \cite{Cur-Phy-72}) that this value should be corrected by a
  multiplicative factor of 1.0020 due to a wrong unit conversion.}
\end{thebibliography}

\end{document}